\documentclass[
superscriptaddress,
amsmath,amssymb,
aps, twocolumn, prb, floatfix
]{revtex4-2}

\usepackage[utf8]{inputenc}
\usepackage[T1]{fontenc}
\usepackage{mathptmx}
\usepackage{tabularx}
\usepackage{booktabs}
\usepackage{array}   
\usepackage[italian,english]{babel}
\usepackage[dvipsnames]{xcolor}

\usepackage[breaklinks,pdftex,pdfpagelabels,bookmarks]{hyperref}
\hypersetup{colorlinks=true,linktocpage=true,pdfstartpage=3,pdfstartview=FitH,breaklinks=true,pdfpagemode=UseNone,pageanchor=true,pdfpagemode=UseOutlines,plainpages=false,bookmarksnumbered, bookmarksopen=true,bookmarksopenlevel=1,hypertexnames=true,urlcolor=Brown,linkcolor=MidnightBlue!60!Black,citecolor=PineGreen}
\usepackage{cleveref}
\usepackage[version=4]{mhchem}
\usepackage{physics}
\usepackage{siunitx}
\usepackage{comment}
\usepackage{graphicx}
\usepackage{dcolumn}
\usepackage{bm}
\usepackage{amsmath}
\usepackage{mathtools, nccmath}
\usepackage{amsfonts}
\usepackage{amssymb}
\usepackage{xcolor}
\usepackage{bbold}

\usepackage{multirow}
\graphicspath{{./Images/}}

\begin{document}
	
\preprint{APS/123-QED}
	
\title{
Supercurrent from the imaginary part of the Andreev levels in non-Hermitian Josephson junctions}
	
\author{R. Capecelatro}
\email{rcapecelatro@unisa.it} 
\affiliation{Dipartimento di Fisica "E.R. Caianiello"$,$ Università di Salerno$,$ via Giovanni Paolo II$,$ 132$,$ I-84084$,$ Fisciano (SA)$,$ Italy}
\author{M. Marciani}
\email{marco.marciani89@gmail.com} 
\affiliation{Dipartimento di Fisica E. Pancini$,$ Università degli Studi di Napoli Federico II$,$ Monte S. Angelo$,$ via Cinthia$,$ I-80126 Napoli$,$ Italy}
\author{G. Campagnano}
\affiliation{CNR-SPIN$,$ UOS Napoli$,$ Monte S. Angelo$,$ via Cinthia$,$ I-80126 Napoli$,$ Italy}
\author{R. Citro}
\affiliation{Dipartimento di Fisica "E.R. Caianiello"$,$ Università di Salerno$,$ via Giovanni Paolo II$,$ 132$,$ I-84084$,$ Fisciano (SA)$,$ Italy}
\affiliation{INFN$,$ Sezione di Napoli$,$ Gruppo collegato di Salerno$,$ Italy}
\author{P. Lucignano}
\affiliation{Dipartimento di Fisica E. Pancini$,$ Università degli Studi di Napoli Federico II$,$ Monte S. Angelo$,$ via Cinthia$,$ I-80126 Napoli$,$ Italy}

\begin{abstract}
We investigate the electronic transport properties of a superconductor–quantum dot–superconductor Josephson junction coupled to a ferromagnetic metal reservoir in the presence of an external magnetic field. The device is described by an effective non-Hermitian Hamiltonian, whose complex eigenvalues encode the energy (real part) and the broadening (imaginary part) of the Andreev quasi-bound states. When extending the Andreev current formula to the non-Hermitian case, a novel contribution arises that is proportional to the phase derivative of the levels broadening. This term becomes particularly relevant in the presence of exceptional points (EPs) in the spectrum, but its experimental detection is not straightforward.
We identify optimal Andreev spectrum configurations where this novel current contribution can be clearly highlighted, and we outline an experimental protocol for its detection. 
We point out that the phase dependence in the levels imaginary part originates from the breaking of a time-reversal-like symmetry. In particular, spectral configurations in the broken phase of the symmetry and without EPs can be obtained, where this novel contribution can be easily resolved.
The proposed protocol would allow to probe for the first time a fingerprint of non-Hermiticity in open junctions that is not strictly related to the presence of EPs.
\end{abstract}
\maketitle

\section{Introduction}

Condensed matter systems described by non-Hermitian (NH) Hamiltonians~\cite{Philip2018, Chen2018, Bergholtz2019, Budich2020, Cayao2023_2, Mi14, SanJose2016, Avila:2019, Cayao2023, Arouca2023, Sayyad2023, Cayao2024, CayaoAguado2024, Kawabata2018, Cayao2022, Kornich2022, Kornich2022_2, Kornich2023, Kokhanchik2023, Paya2025} are increasingly attracting interest, in view of exploring the peculiar physics of open quantum systems~\cite{BreuerPetruccione, Weiss, Gong_2020, Bender07, Bender:1998, Kawabata2019} already observed in optics~\cite{Berry2004, Pen14, Doppler2016, St-Jean2017,Zhou2018,El-Ganainy2019,schonleber2016,Xu2016}.
In this context, using electronic transport measurements to probe NH phenomena, such as exceptional points (EPs) in the Hamiltonian spectrum~\cite{Zhen2015, Shen2018, Cerjan2019, Bergholtz2021, Oku23}, where eigenvectors and eigenvalues coalesce, and NH skin effect~\cite{FoaTorres2018, Yao18, Zhang2020, Zhang:2022, ZhangKai2022}, would pave the way for accessing novel NH topological states and simulating NH dynamics directly within solid state architectures~\cite{Mandal2020, Zhang:2021, Geng2023, Qing2024}.

Recent research has focused on superconducting Josephson junctions (JJs) coupled to normal-metal reservoirs, where the subgap Andreev levels provide a suitable platform for studying NH phenomena in discrete-level systems~\cite{Shen2024, Li2024, Beenakker2024,  CayaoSato2024, CayaoSato2024_2, Capecelatro2025, Pino2025, Ohnmacht2025, Solow2025, Junjie2025, Li2025, CayaoSato2026, Sten2025}. In equilibrium conditions, the interaction with the bath is accounted for by an imaginary self-energy term in the Green's function of the junction~\cite{Shen2024, CayaoSato2024, Pino2025} or of the barrier alone~\cite{Capecelatro2025, Solow2025}, from which an effective NH Hamiltonian with complex eigenvalues can be derived.
The imaginary part of these eigenvalues represents the energy broadening of the so-called Andreev quasi-bound states (quasi-ABS) of the system. Having a complex energy spectrum raises the question of how the quasi-ABS are related to the supercurrent, as well as to other thermodynamic observables~\cite{Li2024, Beenakker2024, CayaoSato2024, CayaoSato2024_2, Capecelatro2025, Pino2025}.
In particular, an open issue revolves around the possibility to express the Josephson current as the phase derivative of the Andreev levels, namely $J\propto\partial_\phi\sum_{\varepsilon_{A}<0}\varepsilon_{A}(\phi)$~\cite{Beenakker1992}, in NH systems.
A straightforward extension of the formula for Hermitian JJs only involving the real part of the levels has been intensively discussed~\cite{Shen2024, Li2024, CayaoSato2024,CayaoSato2024_2, Beenakker2024, Capecelatro2025, Pino2025, Solow2025}.

\begin{figure*}[hbt!]
\centering
\includegraphics[scale=0.29]{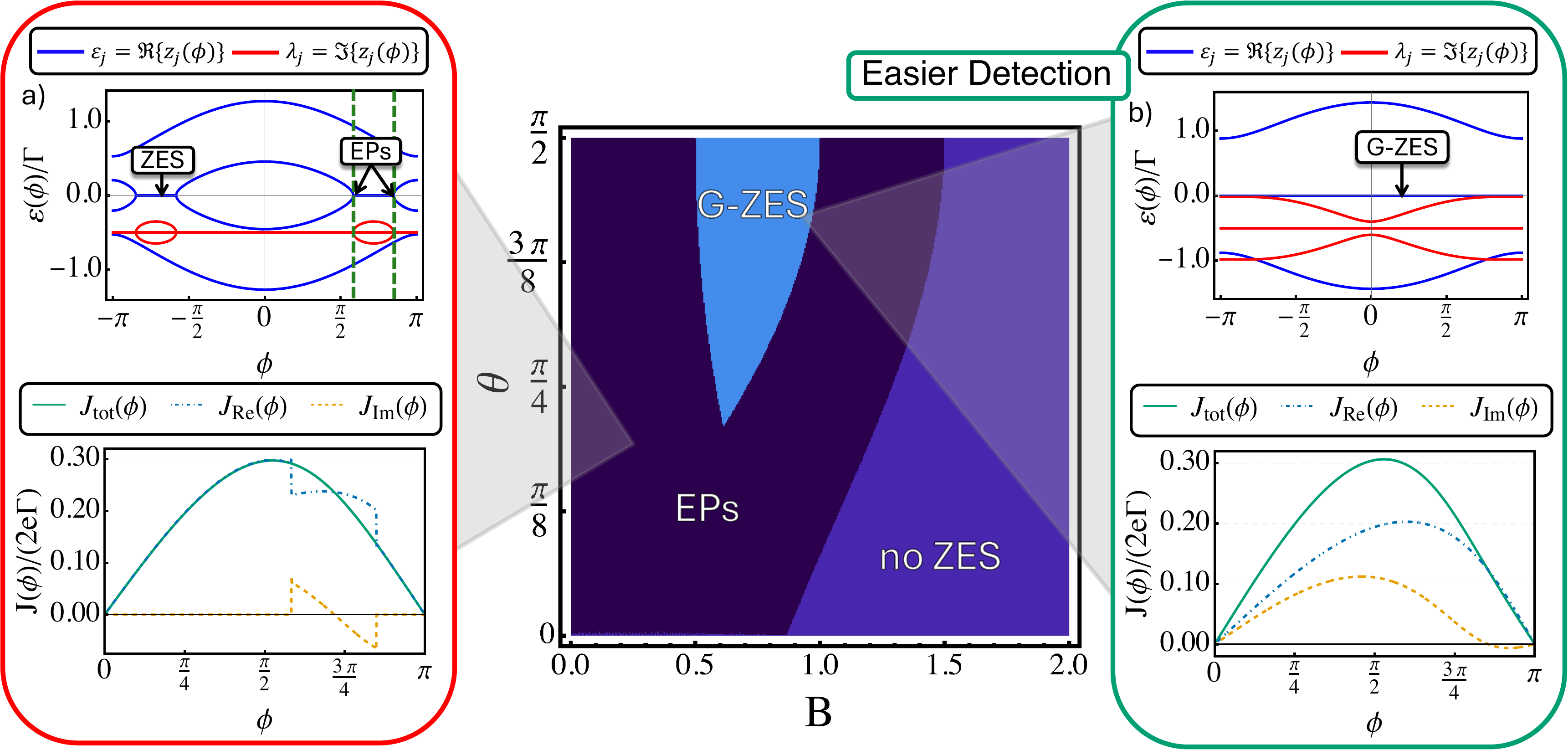}
\caption{
Regions of the parameters space, magnetic field amplitude $B$ (x-axis) and its angle $\theta$ with the ferromagnet magnetization (z-axis), hosting exceptional points (EPs), one pair of global zero-energy states (G-ZES) and without ZES (noZES). In the insets, the corresponding Andreev spectra along with the corresponding CPR. Here, we also show the CPR components coming from the phase dispersion in the real and imaginary parts of the levels, i.e. $J_{\Re}$ and $J_{\Im}$.
Between two zero-energy EPs (ZE-EPs), e.g. phase interval among the green-dashed lines, the Andreev levels are pinned at zero energy giving rise to ZES.
For a JJ hosting EPs (a) $J_{\Im}$ is piecewise defined while in the G-ZES regime (b) it is continue and its detection should be easier.
}
\label{fig: 1_EPs_in_ABS_spectrum}
\end{figure*}

In recent works, the role of the imaginary part of the levels in the supercurrent calculation has been investigated with novel NH Hamiltonian approaches derived from scattering matrix and Green's function (GF) techniques~\cite{Beenakker2024, Capecelatro2025, Pino2025}. 
The quasi-ABS broadening has been predicted to both dampen the Andreev levels current~\cite{Beenakker2024} and, notably, to yield a further contribution due to its phase dispersion~\cite{Capecelatro2025, Pino2025}, i.e. proportional to its derivative with respect to the phase bias. Such contribution becomes relevant when searching in the current-phase relation (CPR) for fingerprints of EPs arising in the ABS spectrum. At these spectral points the energies of two quasi-ABS coalesce, closing a real gap and subsequently opening an imaginary one. Specifically, the real part of the levels merge with a square-root behavior, showing a divergent phase-derivative, and their imaginary part bifurcate with analogous trend.
For this reason, in the presence of EPs, a supercurrent formula accounting solely for the real part of the eigenvalues~\cite{Li2024, CayaoSato2024, CayaoSato2024_2}, 
predicts spikes in the CPR~\cite{CayaoSato2024, CayaoSato2024_2}. However, they do not appear when utilizing exact transport formalism~\cite{Zagoskin, Furusaki1994, Cuevas1996, Solow2025, Meng2009_PRB, Zazunov2009, JonckhereeMartin2009, Asano2019, Minutillo2021} nor the recent approaches based on the NH Hamiltonian~\cite{Shen2024, Pino2025}.
Indeed, the phase dispersion in the imaginary part of the levels provides a finite supercurrent in the interval between two EPs~\cite{Pino2025} that, together with the aforementioned smoothening factor, ensures the continuity of the CPR, thus canceling these spikes.
Therefore, disregarding this CPR component due to the dispersion in the imaginary part can lead to unphysical predictions in the supercurrent behavior.
Yet a protocol for its detection has not been proposed so far.

A promising route is to infer the supercurrent directly from the Andreev levels of the system~\cite{Basset2014, Delagrange2016, Nichele2020}. This approach has become experimentally accessible through recent advances in Andreev spectroscopy, both in normal metal–superconductor heterostructures~\cite{Dirks2011, Lee2014, vanDriel2023} and in SQUID architectures~\cite{Pillet2010, Bretheau2013, Bretheau2013nature, Bretheau2017, Bargerbos2023, Wesdorp2024}. The authors of Ref.~\cite{Nichele2020} reconstructed the Andreev supercurrent from $dI/dV(\phi)$ spectroscopy and benchmarked it against standard CPR measurements, showing excellent agreement. Taken together, these experimental developments suggest that a combination of conductance~\cite{Pillet2010} and CPR measurements~\cite{DellaRocca2007} might be exploited in the detection protocol for the current associated with the imaginary part of the quasi-ABS.

In this work we aim at identifying a system with optimal Andreev spectrum configurations where the novel NH current contribution would be most clearly highlighted in an experiment.
In this respect, we investigate the transport in NH JJs hosting EPs by using the NH current formula that we derived in Ref.~\cite{Capecelatro2025}, which allows us to resolve the imaginary dispersion contribution to the CPR. Specifically, we analyze the underlying physics at the origin of this contribution and search for the key ingredients to enhance it.
Furthermore, we build on the idea in Ref.~\cite{Solow2025} of probing the zero-energy EPs in the ABS spectrum through $dI/dV(\phi)$ spectroscopy and propose a protocol to access the supercurrent due to the levels broadening. In particular, we envision that, in the presence of zero-energy states, this novel current can be detected from the discrepancy between the CPR and the supercurrent reconstructed from $dI/dV(\phi)$. 

We adopt the setup of Ref.~\cite{Solow2025}, a superconductor–quantum dot–superconductor JJ with a ferromagnetic lead coupled to the barrier (SQDFS), under a magnetic field applied to the dot.
The system proves to be promising for the engineering of the Andreev spectrum and the CPR since it is highly tunable and displays zero-energy EPs (ZE-EPs) as soon as the field is non-collinear with the magnetization in F.
The position of the EPs can be found analytically, enabling their exact control via tuning of the system parameter. In this way the phase-interval between two EPs displaying zero-energy states (ZES) with a phase-dispersive imaginary part, Fig.~\ref{fig: 1_EPs_in_ABS_spectrum} (a), can be expanded. Crucially, the interval can be extended to include all phase biases. These are regimes characterized by \emph{globally} extended ZES (global zero-energy states, G-ZES), Fig.~\ref{fig: 1_EPs_in_ABS_spectrum} (b), and are best spots, by design, to perform cross-check measurements of the imaginary-dispersion contribution to the CPR.

To better understand what determines the phase dependence of the levels imaginary part, we investigate the symmetries of the dot NH Hamiltonian. We discover that an additional Time-Reversal-like symmetry (TRS) emerges when subtracting the average losses from the Hamiltonian. We link the emergence of such phase dispersion to a transition between regions where this symmetry holds and those where it is broken. Analogously to what happens in $\mathcal{PT}$-symmetric systems, at the EPs the system enters the broken phase of the TRS and the quasi-ABS imaginary part becomes phase dispersive. 
The symmetry is explicitly broken when dot is tuned out of resonance with respect to the leads chemical potential.
In the off-resonance regime, all the Andreev levels exhibit a phase-dispersive imaginary part, thus providing enhanced visibility of its supercurrent contribution.
All together, our findings shed new light on the detection of pure non-Hermitian effect by the means of standard transport measurements.

The manuscript is organized as follows.
In Sec.~\ref{sec: 2_Model_full_Hamiltonian} we introduce the system under study, Fig.~\ref{fig: 1_SQDFS_Gamma}, and we show how to derive an effective NH Hamiltonian for the QD in Sec.~\ref{sec: 2_Model_GF}. Then, in Sec.~\ref{sec: 2_Model_Josephson_current} we introduce the current formula we use to compute the junction CPR from the complex Andreev levels of the system.
In Sec.~\ref{sec: 3_Asymmetric_JJ_in_Res} we analyze the simple case of an asymmetric QD JJ in resonant tunneling regime, where analytic expressions for the quasi-ABS and the EPs position can be determined so as to be able to tune the system in the G-ZES regimes.
In Sec.~\ref{sec: 4_symmetries} we discuss how the phase dependence of the Andreev levels broadening is related to the Hamiltonian symmetries.
We provide the extension of the analysis to the off-resonance regime in App.~\ref{sec: 5_Off_res_JJ}.
Finally, we outline the conclusions and the outlooks of the work in Sec.~\ref{sec: 6_Conclusions}. 

\section{Model}

\begin{figure}[h!]
	\centering
	\includegraphics[scale=0.151]{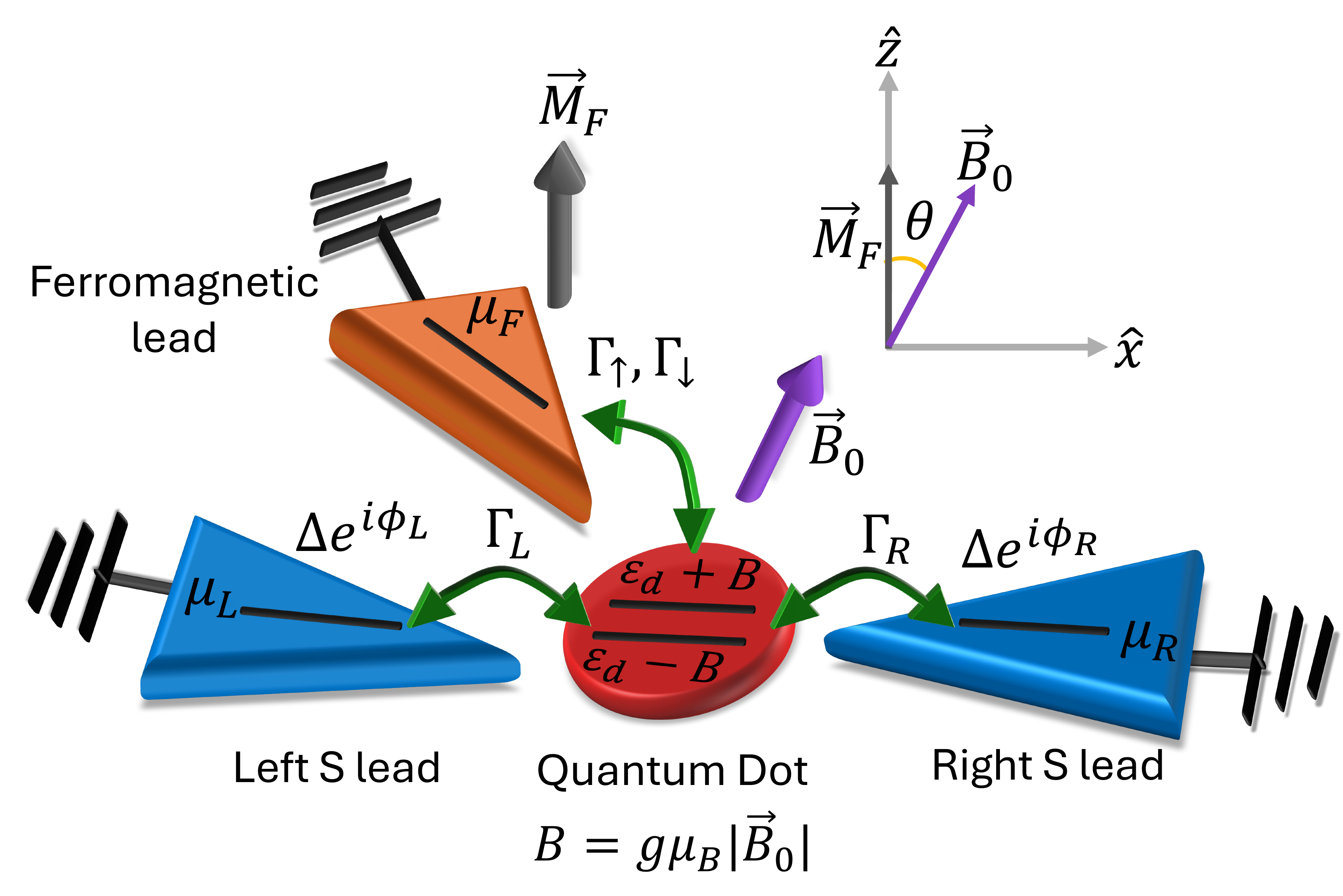}
    \caption{Scheme of the Quantum Dot Josephson junction (QD JJ) connected to the ferromagnetic lead. We show the relevant parameters: the dot energy level $\varepsilon_{d}$; the superconducting gap $\Delta$ and the phase in the left/right lead $\phi_{L/R}$; the chemical potential $\mu_{L/R/F}$ respectively of the left/right superconductors and ferromagnet; the F lead magnetization $\vec{M}_{F}$; the external magnetic field $\vec{B}_{0}$ inducing a Zeeman splitting of order $B=g\mu_{B}|\vec{B}_{0}|$ on the dot;
    the hybridization parameter of the dot with the left/right S leads $\Gamma_{L/R}$ and the spin-dependent hybridizations with the F lead $\Gamma_{\uparrow/\downarrow}$.}
	\label{fig: 1_SQDFS_Gamma}
\end{figure}

In this section we derive the effective NH Hamiltonian for the QD JJ, Fig.~\ref{fig: 1_SQDFS_Gamma}, within Green's function (GF) formalism.
By tracing out the degrees of freedom of the superconductors and of the ferromagnet the \emph{interacting} GF of the QD is obtained. From this we can extract an effective NH Hamiltonian for the barrier only, whose eigenvalues are the complex Andreev levels of the JJ.

\subsection{Full system Hamiltonian}
\label{sec: 2_Model_full_Hamiltonian}
The system under consideration is a JJ with a single level quantum dot (QD) barrier coupled to a ferromagnetic metal lead (F) in the presence of an external magnetic field, Fig.~\ref{fig: 1_SQDFS_Gamma}.
The total Hamiltonian of the system reads as 
\begin{equation}
		\label{Hamiltonian_system}
		H_{\rm sys}=H_{d}+ H_{\rm S } + H_{\rm F}+ H_{\rm T}\,,
\end{equation}
where $H_{d}$, $H_{\rm S }$, $H_{\rm F}$, are the dot, the S leads, and the F lead Hamiltonians, respectively, and $H_{\rm T}$ is the tunneling Hamiltonian between the leads and the dot.
	
The dot Hamiltonian is 
\begin{equation}
    \label{H_d}
    H_{d}=\sum_{\sigma,\sigma'=\uparrow,\downarrow}d_{\sigma}^{\dagger}\left(\varepsilon_d\delta_{\sigma,\sigma'}+\vec{B}\cdot\vec{\sigma}\right) d_{\sigma'},
\end{equation}
with $d_{\sigma}$ ($d_{\sigma}^{\dagger}$) the annihilation (creation) operator on the dot with spin $\sigma$, $\varepsilon_d$ the dot energy and $\vec{\sigma}=(\sigma_{x},\sigma_{y},\sigma_{z})$ the vector of Pauli operators in the spin space. Here, we assume that the magnetic field $\vec{B}_{0}$ induces only a Zeeman field $\vec{B}=g\mu_{B}\vec{B}_0$ on the dot. 

The s-wave superconductors share the same superconducting gap $\Delta$ and are set at the same chemical potential $\mu=\mu_{L}=\mu_{R}$. 
Their Hamiltonian is:
\begin{equation}
		\begin{split}
			H_{\rm S \,}=& \sum_{i=L,R}\sum_{\vec{k}}(\varepsilon_{\vec{k}}-\mu) \left(c_{i,\vec{k},\uparrow}^{\dagger}c_{i,\vec{k},\uparrow} + c_{i,\vec{k},\downarrow}^{\dagger}c_{i,\vec{k},\downarrow}\right) \\
            &+ \Delta e^{i\phi_{i}} c_{i,\vec{k},\uparrow}^{\dagger}c_{i,-\vec{k},\downarrow}^{\dagger} +\rm H.c. \hspace{0.3 mm}
		\end{split}
		\label{H_leads}
\end{equation}
where $c_{i,\vec{k},\sigma}$ operator annihilates electrons in the state $\vec{k}$ with spin $\sigma$ on the lead $i$ ($i=L,R$). Here, $\varepsilon_{ \vec{k}}$ is the normal-state dispersion in momentum space and $\phi_{R/L} = \pm \phi/2$ is the superconducting phase in the lead $R/L$. 

The ferromagnetic reservoir is modeled as a 1D semi-infinite chain with a Zeeman interaction term, induced by the magnetization, $\vec{M}_{F}$, of the ferromagnet.
Its tight-binding Hamiltonian $H_{\rm F}$ reads:
\begin{equation}
        H_{\rm F}=\sum_{\vec{k}}\sum_{\sigma,\sigma'}c_{F,\vec{k},\sigma}^{\dagger}\left((\varepsilon_{F,\,\vec{k}} -\mu_F)\delta_{\sigma,\sigma'}+\vec{M}_F\cdot\vec{\sigma} \right) c_{F,\vec{k},\sigma'}\,,
        \label{H_N_lead}
\end{equation}
with $\mu_{F}$ and $\varepsilon_{F,\, \vec{k}}$ being the chemical potential and the dispersion law, respectively. 
In general, $\vec{M}_{F}$ is not collinear with the magnetic field of the dot. In the following, we choose $z$ as the spin quantization axis of F and $\vec{B}=(B_{x}, 0, B_{z})=(B\sin{\theta}, 0, B\cos{\theta})$, where, without loss of generality we can set $B_{y}=0$.

The tunneling Hamiltonian $H_{\rm T}$ describes the interaction between the dot and both the superconductors and the ferromagnet, thus reading
 \begin{equation}
		H_{\rm T} =\sum_{i=L,R,F} H_{T_{i}}= \sum_{i=L,R,F}t_{i} \sum_{\vec{k}}\sum_{\sigma}  c_{i,\vec{k},\sigma}^{\dagger}d_{\sigma} +\rm H.c. \hspace{0.3 mm},
		\label{H_hopping}
\end{equation}
where $t_{i=L,R,F}$ are the hopping amplitudes, assumed to be real and $\vec{k}$-independent.

At equilibrium we set the chemical potentials to be aligned, i.e. $\mu=\mu_{F}=0$. 

\subsection{Green's function and effective NH Hamiltonian}
\label{sec: 2_Model_GF}
The dot Matsubara GF can be derived exactly by tracing out the degrees of freedom of the left and right S leads and the F bath. Their legacy are self-energy terms, $\Sigma_{S}\left(\omega_{n}\right)=\Sigma_{L}+\Sigma_{R}$ and $\Sigma_{F}(\omega_{n})$, with $\omega_{n}=\pi(2n+1)T$ being the fermionic Matsubara frequencies and $T$ the temperature. 
The dot GF is therefore defined as $G_{d}\left(\omega_{n}\right)=\left(i\omega_{n}- H_{d}-\Sigma_{S}(\omega_{n}) - \Sigma_{F}(\omega_{n})\right)^{-1}$.
Here, 
$\Sigma_{i=L,R,F}(i\omega_{n})=H_{T_{i}}^{\dagger}G^{0}_{i}H_{T_{i}}$
where $G^{0}_{i}$ is the bare GF of the lead $i=L,R,F$ (see Appendices~\ref{app: GF_ferromagnetic_lead}-\ref{app: GF_finite_Delta_case}). 

Representing the dot field operator in the Nambu-spin basis as $\Psi=\left[d_{\uparrow},d_{\downarrow},-d^{\dagger}_{\downarrow}, d^{\dagger}_{\uparrow}\right]$, we can write an explicit expression for the dot GF matrix in Nambu-spin space. 
In the manuscript, we indicate with $\check{.}$ the $4\times4 $ matrices in Nambu$\otimes$Spin space and with $\hat{.}$ the $2\times2$ matrices in Nambu and Spin spaces.

The self-energies of the S leads, $\check{\Sigma}_{L/R}$, have a simple form when we assume that the superconductors are described by a flat conduction band with a constant density of states, $\rho_{0}$, at the Fermi level~\cite{Meng2009_PRB, Zazunov2009,JonckhereeMartin2009,Benjamin2007,Capecelatro2023}. $\check{\Sigma}_{L/R}$ are then proportional to the relative hybridization parameters with the dot, $\Gamma_{L/R}=\pi \rho_{0}t_{L/R}^2$.

In the weak-coupling regime, i.e. for large superconducting gap $\Delta\gg\Gamma_{L,R}$, $\check{\Sigma}_{S}(\omega_n)=\check{\Sigma}_{L}+\check{\Sigma}_{R}$ can be approximated as a frequency independent term, Appendix~\ref{app: GF_finite_Delta_case}.
When introducing the total hybridization, $\Gamma=\Gamma_{L}+\Gamma_{R}$, and the junction asymmetry, $\gamma=\Gamma_{L}-\Gamma_{R}$, this reads
\begin{equation}
    \label{S_leads_self_energy}
    \check \Sigma_{S}= \Gamma\cos{\left(\frac{\phi}{2}\right)}\hat{\tau}_{x}\otimes \hat{1}+\gamma\sin{\left(\frac{\phi}{2}\right)}\hat{\tau}_{y}\otimes \hat{1}\,.  
\end{equation}

Hereafter we denote identity matrices with the symbol $\check{1}$ $(\hat{1})$ and the analogue of Pauli matrices in Nambu space with $\hat{\tau}_{x,y,z}$.

Analogously, the self-energy $\hat{\Sigma}_{F}$ can be written in terms of the hybridizations $\Gamma_{\uparrow,\downarrow}$ for the F lead, where the spin occupation imbalance at the chemical potential yields spin-resolved parameters.

$\hat{\Sigma}_{F}$ is derived in Appendix~\ref{app: GF_ferromagnetic_lead} in the \emph{broadband} approximation for which the bandwidth of the F lead is taken much larger than any other energy scale of the JJ. In this limit, the bath induces only decoherence to the system so that $\Sigma_{F}$ becomes a simple imaginary and frequency-independent term 
\begin{eqnarray}
    \label{F_lead_self_energy_broadband}
    \check{\Sigma}_{F}&=&-\frac{i\,\mathrm{sign}(\omega_{n})}{2}\left[\Gamma_{\uparrow}\left(\check{1}+\hat\tau_{z}\otimes\hat{\sigma_{z}}\right)-\Gamma_{\downarrow}\left(\check{1}-\hat\tau_{z}\otimes\hat{\sigma_{z}}\right)\right]\nonumber\\
    &=&-i\mathrm{sign}(\omega_{n})\Gamma_{N}\check{1}-i\mathrm{sign}(\omega_{n})\gamma_{N}\hat{\tau}_{z}\otimes\hat{\sigma}_{z}\,,
\end{eqnarray}
where $\Gamma_{N}$ is the average dissipation, $\Gamma_{N}=(\Gamma_{\uparrow}+\Gamma_{\downarrow})/2$, and $\gamma_{N}$ is the dissipation imbalance, $\gamma_{N}=(\Gamma_{\uparrow}-\Gamma_{\downarrow})/2$.

In the large $\Delta$ limit (see Appendix~\ref{app: GF_finite_Delta_case}), starting from the retarded dot GF we can identify an effective frequency-independent NH Hamiltonian
\begin{equation}
    \check{G}_{d}(z)=\left[ z\check{1}-\check{H}_{d}-\check{\Sigma}_{S}-\check{\Sigma}_{F}\right]^{-1} \approx\left[ z\check{1}-\check{H}_{eff}\right]^{-1}\,,
\end{equation}
with
\begin{eqnarray}
\label{Heff_NH}
\check{H}_{eff}&=&\varepsilon_{d}\hat{\tau}_{z}\otimes\hat{1}+B_{z}\hat{1}\otimes\hat{\sigma}_{z}+B_{x}\hat{1}\otimes\hat{\sigma}_{x}-i\Gamma_{N}\check{1}-i\gamma_{N}\hat{\tau}_{z}\otimes\hat{\sigma}_{z}\nonumber\\
    &&+\Gamma\cos{\left(\frac{\phi}{2}\right)}\hat{\tau}_{x}\otimes \hat{1}+\gamma\sin{\left(\frac{\phi}{2}\right)}\hat{\tau}_{y}\otimes \hat{1}\,.
\end{eqnarray}

In Hermitian junctions, the weak-coupling condition \cite{Beenakker1992} guarantees that the ABS suffice to describe the Josephson current, while the contribution from the supragap continuum is negligible. 
Crucially, this property has been demonstrated to stay true in open junctions \cite{Beenakker2024, Capecelatro2025} -- where poles of the dot GF or, alternatively, the complex eigenvalues of the NH Hamiltonian are the heritage of the subgap states, called quasi-ABS -- provided that we also have $\Gamma_{N}\ll\Delta$. 
Therefore, this regime is found convenient to compare and benchmark the different CPR formulas for the quasi-ABS directly against exact GF calculations, which in general would naturally embody also the contribution from the continuum.

\subsection{Josephson current from Andreev quasi-bound states}
\label{sec: 2_Model_Josephson_current}
In Ref.~\cite{Capecelatro2025} the following current formula for the quasi-ABS was derived within the exact GF formalism
\begin{eqnarray}
\label{Jpol_general}
J_{ABS}(\phi) =J_{\Re}+J_{\Im}&=&  -\frac{e}{\pi} \sum_{j}\;     \left[(\partial_\phi\varepsilon_{j})\,                    \mathrm{Im}\,L_j - (\partial_\phi\lambda_{j}) \,        \mathrm{Re}\,L_j\right],\nonumber  \\
    L_j &=& \int_{-\Delta}^{\Delta}         \mathrm{d}\omega\; \frac{n_{F}        (\omega)}{\omega - z_{j}} ,
\end{eqnarray}
where $z_{j}=\varepsilon_{j}-i\lambda_{j}$ are the complex energies of the quasi-ABS, $e$ is the electron charge, $n_{F}(\omega)$ is the Fermi function and the summation runs over all the complex eigenvalues.

Such formula points out that two distinct terms contribute to the subgap current, $J_{\Re}$ coming from the dispersion of the real-part of the quasi-ABS and $J_{\Im}$ coming from the dispersion of the imaginary one. Altogether, Eq.~\ref{Jpol_general} provides the chance to predict the weight of the latter contribution to the CPR in a generic system.

We show in Appendix~\ref{app: CPR_finite_infinite_Delta_cases} that, in the zero temperature and large gap limit, $T\rightarrow0$ and $\Delta\rightarrow\infty$, this is further simplified to 
\begin{eqnarray}
	\label{Jpol_T0_simp}
	J_{ABS}(\phi) &\overset{\Delta\rightarrow\infty}{\overset{T\rightarrow0}{=}}&J_{\Re}+J_{\Im}= \sum_{j}J_{\Re, j}+\sum_{j}J_{\Im, j}\;, \;\; \mathrm{with} \nonumber\\
    J_{\Re, j}&\overset{\Delta\rightarrow\infty}{\overset{T\rightarrow0}{=}}&-\frac{e}{\pi} \partial_\phi \varepsilon_j\left( \arctan\left(\varepsilon_j/\lambda_j\right) - \frac{\pi}{2}\right),\\
	J_{\Im, j}&\overset{\Delta\rightarrow\infty}{\overset{T\rightarrow0}{=}}& -\frac{e}{\pi} \;\partial_\phi \lambda_j\ln\left( |z_j|\right)\,\nonumber.
\end{eqnarray}

We observe that Eq.~\ref{Jpol_T0_simp} can be proven to be equivalent to the current equation of Ref.~\cite{Shen2024}, see Appendix~\ref{app: CPR_finite_infinite_Delta_cases}, 
if, instead of the full junction effective Hamiltonian, only that of the barrier (here the QD) is plugged in~\cite{Capecelatro2025}.
Further, in Appendix~\ref{app: CPR_in_TRS_case} we show that in specific conditions we can express the CPR only in terms of the quasi-ABS below the chemical potential, and also recover an expression for $J_{\Im}$ contribution in the regions delimited by EPs.

In the following sections we analyze the situations where the imaginary part of the levels acquires a phase dependence so that $J_{\Im}$ becomes relevant to predict the correct CPR and it can be conveniently detected.

\section{
Global ZES and the optimal regime for the detection of \texorpdfstring{$J_{Im}$}{JIm}}
\label{sec: 3_Asymmetric_JJ_in_Res}

In this section, we investigate how to drive the junction into regimes where the imaginary part of the quasi-ABS is dispersive at all phase biases. 
In particular, we focus on engineering the Andreev levels in order to obtain spectral configurations hosting G-ZES.
These regimes are found to be convenient for highlighting $J_{\Im}$ in future experiments.
We specify our analysis to the resonant tunneling regime, $\varepsilon_{d}=0$.
In this limit, analytic expressions for the complex Andreev levels and for the EPs can be derived, thus allowing a simple understanding of the key ingredients to highlight $J_{\Im}$ in experiments. Our analysis is extended to the non-resonant case in App.~\ref{sec: 5_Off_res_JJ}.

\subsection{Analytical expressions for quasi-ABS and EPs}
\label{sec: 3_analytic_ABS}
 
In this limit, analytic expressions for the complex Andreev levels can be derived.
The quasi-ABS are obtained diagonalizating $\check{H}_{eff}$ in Eq.~\ref{Heff_NH}:
\begin{eqnarray}
\label{Andreev_levels} 
z_{j=1,\ldots,4}
&=& -\,i\,\Gamma_N \;\pm\; 
\sqrt{\,B^{2} + \alpha^2(\phi) - \gamma_N^{2}
   \;\;\pm\; 2\,\beta^2(\phi,\theta)}\,\, \nonumber\\[5pt]
   &&\alpha(\phi) 
= \sqrt{\sin^{2}\!\left(\tfrac{\phi}{2}\right)\gamma^{2}
     + \cos^{2}\!\left(\tfrac{\phi}{2}\right)\Gamma^{2}} \\[5pt]
 &&\beta(\phi,\theta)= \sqrt[4]{\,B^{2}\Big(\alpha^2(\phi) - \gamma_N^{2}\cos^{2}\theta\Big)}.\nonumber\,
\end{eqnarray}

\begin{figure}[h!]
    \centering
    \includegraphics[scale=0.36]{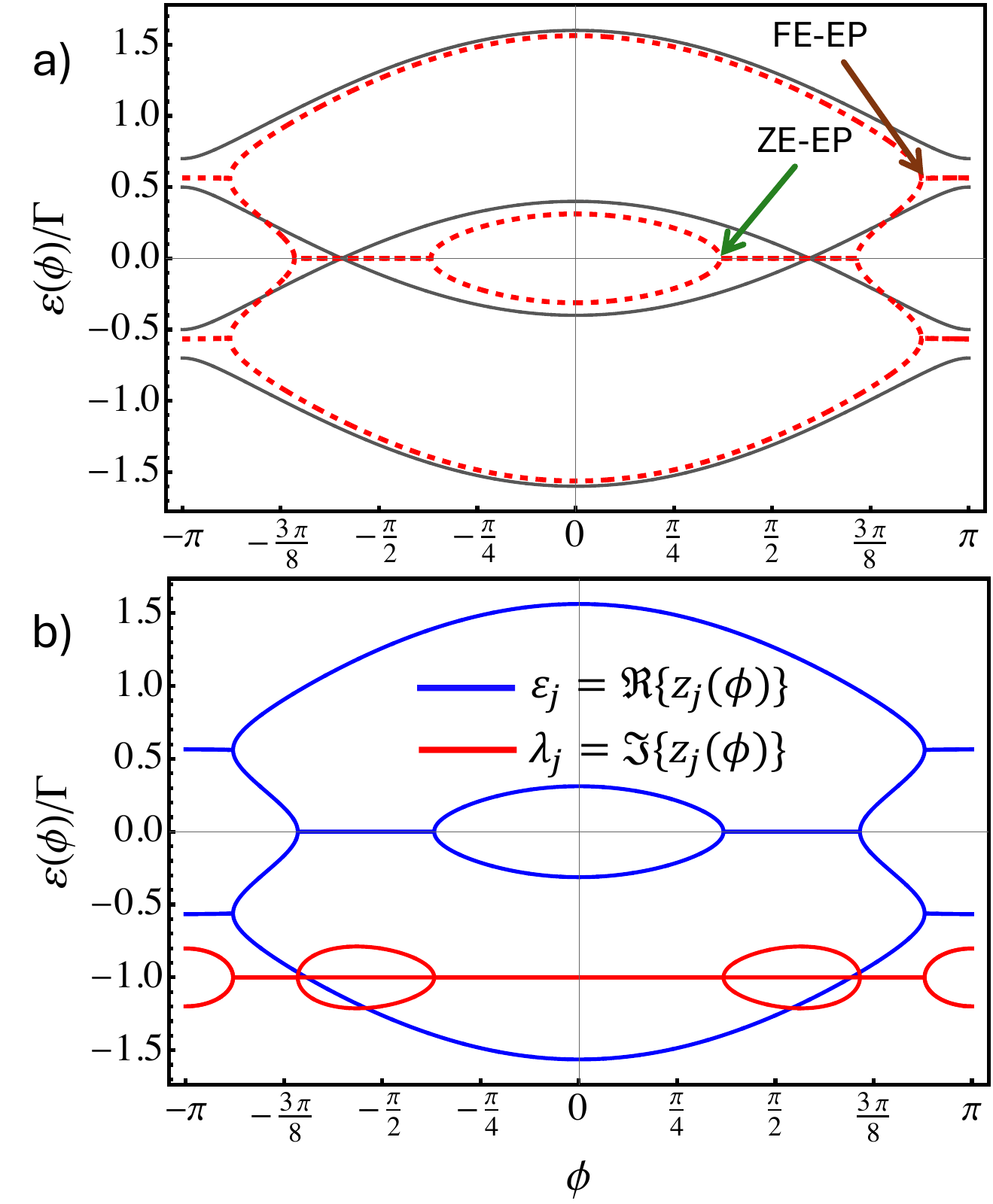}
    \caption{Schematic representation of the emergence of the EPs in a NH JJ at $\varepsilon_{d}=0$ (a), along with the complex Andreev spectrum (b). In (a) ABS spectrum of the Hermitian system, i.e. $\Gamma_{N}=0$ (gray lines), and the real part of the quasi-ABS of its NH counterpart (red-dashed lines). ZE-EPs and FE-EPs are shown.}
    \label{fig: 4_Prototype_spectrum}
\end{figure}

The structure of the spectrum is shaped by the adjoint particle-hole symmetry, PHS$^{\dagger}$, stemming from the superconductivity~\cite{Solow2025,CayaoSato2024,Capecelatro2025,Pino2025}. This is fulfilled both by pairs $(z_j,-z_j^*)$
or by unpaired levels with $\Re\{z_j\}=0$ and allows in principle for the appearance of EPs at both finite and zero energy, i.e. FE-EPs and ZE-EPs.
An example of a typical spectral configuration for the ABS is shown in Fig.~\ref{fig: 4_Prototype_spectrum}.

We note that the average coupling to the normal metal lead, i.e. $\Gamma_{N}$, gives only an imaginary shift of the quasi-ABS. Therefore, the phase dispersion emerges in the imaginary part of the levels only when the radicand of the square-root in Eq.~\ref{Andreev_levels} becomes either real negative or complex, which happens respectively in the regions enclosed by ZE-EPs and those extending beyond the FE-EPs.
This property of the Andreev spectrum cannot be explained solely in terms of the PHS$^{\dagger}$ but, importantly, it requires the consideration an additional TRS-like symmetry that we describe in Sec.~\ref{sec: 4_symmetries}.

Physically, we observe that the FE-EPs arise from the coalescence of two Kramer's partner levels, i.e. spin-up electron-like excitation with spin-up hole-like one, Fig.~\ref{fig: 4_Prototype_spectrum} (a). These can be present when there is an imbalance between the two spin channels in the QD-F couplings, i.e. $\Gamma_{\uparrow}\neq\Gamma_{\downarrow}$.
Contrary, the presence of ZE-EPs is only unlocked by the non-collinearity between $\vec{M}_{F}$ and $\vec{B}$. This gives rise to spin-mixing terms in $\check{H}_{eff}$ that couple two particle-hole symmetric quasi-ABS, with opposite spin, involved in the zero-energy crossings, see the gray lines in Fig.~\ref{fig: 4_Prototype_spectrum} (a). These levels thus evolve into ZES delimited by ZE-EPs.

The position of these EPs are analytically determined: 
\begin{eqnarray}
\label{EPs_position}
\phi_{\text{FE-EP}}^{ext} 
&=& \pm\arccos\left( 
\frac{2\gamma_N^{2} \cos^{2}\theta-\gamma^{2} - \Gamma^{2} }
{\Gamma^{2}-\gamma^{2}}
\right),\\[1em]
\phi_{\text{ZE-EP}}^{ext} 
&=&\pm\arccos\left( 
\frac{ 2 \gamma_N^{2}+2 B^{2} - 4 B  \gamma_N \sin \theta -\gamma^{2} - \Gamma^{2}}
{\Gamma^{2}-\gamma^{2}}
\right), \nonumber \\[1em]
\phi_{\text{ZE-EP}}^{int} 
&=& \pm\arccos\left( 
\frac{2\gamma_N^{2} + 2B^{2} + 4 B \gamma_N \sin \theta-\gamma^{2} - \Gamma^{2} }
{\Gamma^{2}-\gamma^{2}}
\right) \,,\nonumber 
\end{eqnarray}
where the labels \emph{int} (\emph{ext}) refer to the ZE-EP closer to $\phi=0$ ($\phi=\pi$).

We note that, for $\theta=0$, 
$\phi_{\text{ZE-EP}}^{ext}=\phi_{\text{ZE-EP}}^{int}$ and thus the EPs reduce to simple crossings between two Andreev levels, e.g. gray lines in Fig.~\ref{fig: 4_Prototype_spectrum} (a).

In this spectral configuration, the presence of FE-EPs prevents the formation of zero-energy states that extend to $\phi=\pi$. Indeed, the ZE-EPs are prevented by the PHS$^\dagger$ in the regions after the FE-EPs, where the levels are paired at the same finite energy, Fig.~\ref{fig: 4_Prototype_spectrum} (b).

Nonetheless, the phase of all the EPs can be controlled by tuning the junction asymmetry, $\gamma=(\Gamma_L-\Gamma_R)/2$, the dissipation imbalance between the two spin channels, $\gamma_{N}$, and the magnetic field, $\vec{B}$. 
Interestingly, having different couplings between the QD and the two S leads allows to build globally extended ZES and quasi-ABS that have a phase-dispersive imaginary part at all phase biases. 

\begin{figure}[h!]
    \centering
    \includegraphics[scale=0.52]{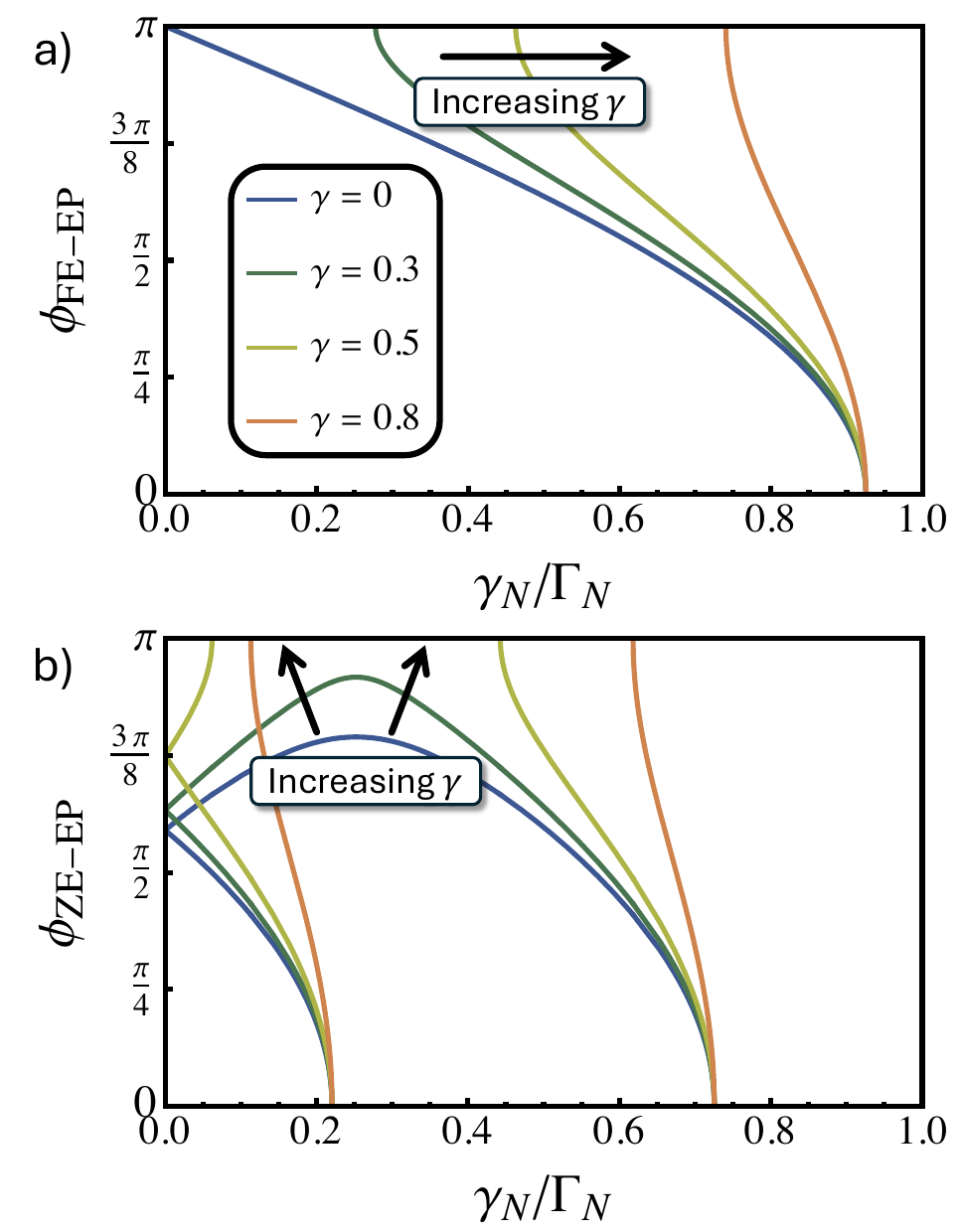}
    \caption{The position of the external FE-EP (a) and the ZE-EPs (b) vs. the dissipation imbalance for the two spins in F, $\gamma_N/\Gamma_N$, for different values of the coupling asymmetry $\gamma$. The more asymmetric is the junction the higher is the $\gamma_N$ value at which FE-EPs manifest.
    External ZE-EPs, in the absence of finite energy ones, can annihilate in pairs at $\phi=\pm\pi$, while internal ones tend to annihilate at $\phi=0$. The system parameters are $\Gamma=1$, $\Gamma_N=2$, $B=0.6$, $\theta=1$.}
    \label{fig: 14 External_EPs_diff_gamma}
\end{figure}

Specifically, we observe that the formation of FE-EPs, which emerge as soon as $\gamma_N\neq0$ for symmetric JJs ($\gamma=0$), can be retarded by setting $\gamma\neq0$, see Fig.~\ref{fig: 14 External_EPs_diff_gamma} (a) which shows the phase position of the FE-EP as a function of $\gamma_{N}/\Gamma_{N}$. 

\begin{figure*}
    \centering
    \includegraphics[scale=0.365]{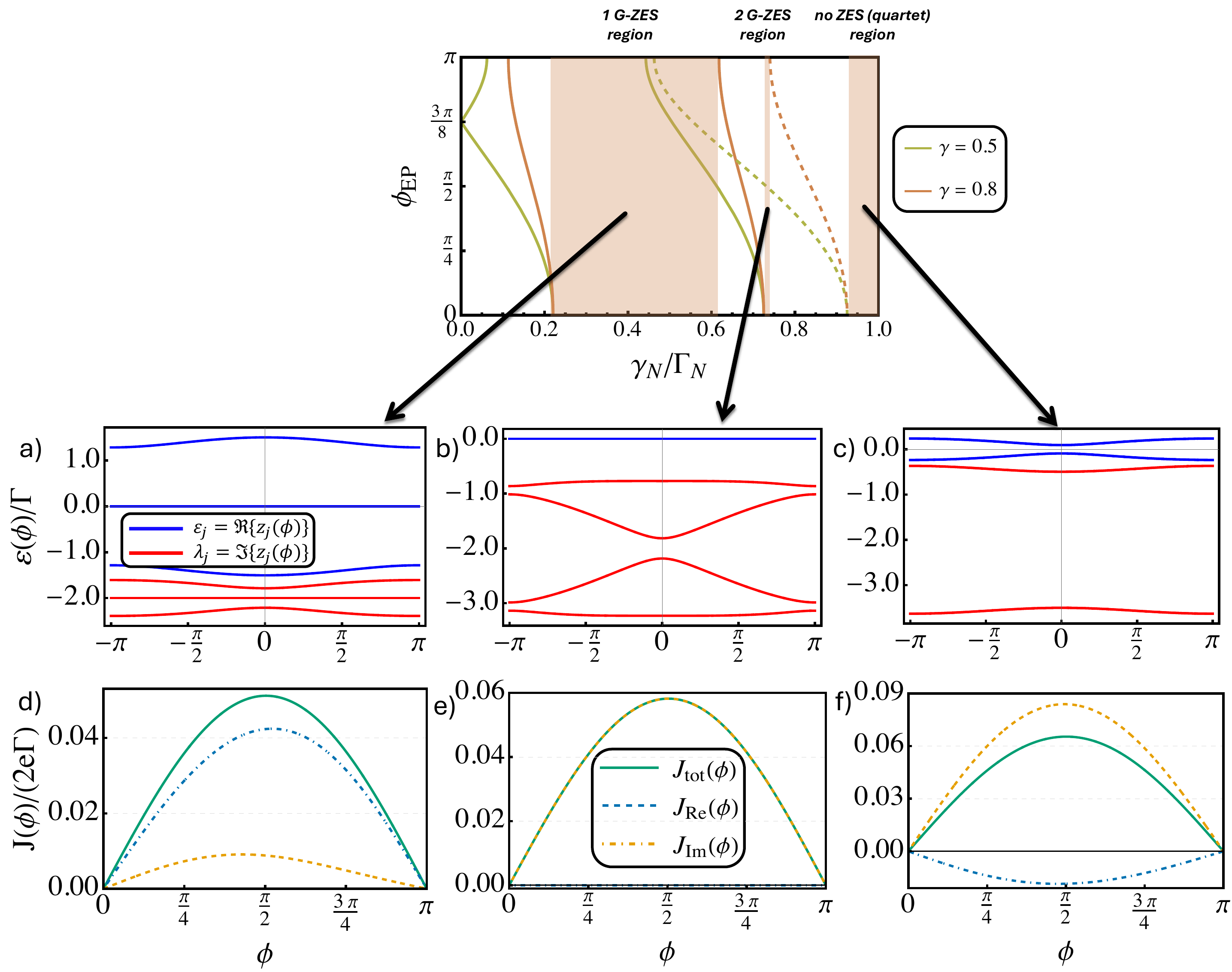}
    \caption{Emergence of regions with spectral configurations without EPs and with phase-dependent level broadening, along with the corresponding CPRs. In the upper panel: phase positions of all EPs, ZE-EPs ( thick lines) and FE-EPs  (dashed lines) vs. $\gamma_N/\Gamma_N$ for different values of the S-leads coupling asymmetry, $\gamma=0.5,\,0.8$. 
    The $\gamma_N$-windows with no EPs in the spectrum for the JJ with $\gamma=0.8$ are shaded in brown. 
    The system parameters are the same of Fig.~\ref{fig: 14 External_EPs_diff_gamma}.
    In the lower panels: typical spectra in each of the three shaded regions of the parameters space, 1G-ZES (a) 2G-ZES (b) and no ZES quartet (c), along with their corresponding CPR contributions, $J_{\Re}$ and $J_{\Im}$ (d-f).
    Here all parameters stay the same, save $\gamma=0.8$ and $\gamma_N/\Gamma_{N}=0.25,\,0.73,\,0.95$ respectively in (a,d), (b,e) and (c,f).}
    \label{fig: 5_Panel_ABS_current_Asymm_JJ}
\end{figure*}

In this sense, the JJ asymmetry makes the junction resilient to host this kind of EPs. 
At the same time we note that increasing $\gamma$ enlarges the separation between the two ZE-EPs thus leading to an expansion of the ZES lines in between. This effect is particularly pronounced when the magnetic field is strongly tilted with respect to the F lead magnetization, i.e. $\theta\gtrsim\pi/4$, see the $\gamma=0.3$ curve in Fig.~\ref{fig: 14 External_EPs_diff_gamma} (b).

A peculiar consequence of these facts is the following.
When the asymmetry of the JJ increases, e.g. $\gamma=0.5$ in Fig.~\ref{fig: 14 External_EPs_diff_gamma} (a), and the emergence of FE-EPs in the spectrum is shifted at high $\gamma_{N}$ values, the external ZE-EPs are free to move toward $\phi=\pm\pi$ and exit the phase interval $\phi\in[-\pi,\pi]$, Fig.~\ref{fig: 14 External_EPs_diff_gamma} (b).
In other words, when FE-EPs are absent, ZE-EPs can annihilate at $\phi=±\pi$ with those coming from the outer phase intervals, i.e. $\phi\in[\pm\pi,\pm3\pi]$.
At the same time, increasing $\gamma_{N}$ drives the internal EPs toward annihilation at $\phi=0$, Fig.~\ref{fig: 14 External_EPs_diff_gamma} (b).
The external ZE-EPs then re-appear in the phase range $[-\pi,\pi]$ at higher $\gamma_{N}$ values before undergoing annihilation at $\phi=0$ too.
An even more extreme situation is found for very large $\gamma$ where the two ZE-EPs disentangle and appear in the spectrum independently, e.g. $\gamma=0.8$ in Fig.~\ref{fig: 14 External_EPs_diff_gamma} (b).
In those parameters regions where all the ZE-EPs have been annihilated pairwise, e.g. $\gamma=0.5,\,0.8$ in Fig.~\ref{fig: 14 External_EPs_diff_gamma} (b), globally extended ZES become accessible.

\subsection{Andreev spectrum configurations with phase-dependent levels broadenings}
\label{sec: 3_spectral_regimes_GZES}

Combining the information about FE- and ZE-EPs positions, we can easily observe the entire EPs dynamics of the JJ and explain the evolution of ZES enclosed by ZE-EPs into global zero-energy states (G-ZES), living at all phase biases.
In order to access and analyze all possible regimes where the quasi-ABS have phase-dependent imaginary parts, we study the case of high dissipation $\Gamma_{N}=2 \Gamma$. Nonetheless, the most relevant regimes, those featuring one pair of G-ZES in the spectrum, are realized even at much lower dissipation where the supercurrent stays sizable, e.g. Fig.~\ref{fig: 1_EPs_in_ABS_spectrum} (b) obtained for $\Gamma_{N}=\gamma_{N}=0.5\Gamma$, see also Sec.~\ref{sec: 3_measurability}. 

\subsubsection{1G-ZES}

For $\gamma=0.5,\,0.8$,  we can get some phase windows with no EPs in the Andreev spectrum, upper panel of Fig.~\ref{fig: 5_Panel_ABS_current_Asymm_JJ}. 
Specifically, the appearance of FE-EPs, dashed lines in upper panel of Fig.~\ref{fig: 5_Panel_ABS_current_Asymm_JJ}, is strongly retarded so that ZE-EPs, thick lines, can annihilate in pairs at $\phi=0$ and $\phi=\pm\pi$.
In the $\gamma_N$ interval between the annihilation of internal ZE-EP and the reappearance of the external one the quasi-ABS spectrum host a pair of G-ZES, e.g. the  \emph{1 G-ZES} region in Fig.~\ref{fig: 5_Panel_ABS_current_Asymm_JJ}, that is the brown-shaded region between $\gamma=0.8$ thick lines.

The G-ZES carry supercurrent only via the phase-dispersive imaginary part, Fig.~\ref{fig: 5_Panel_ABS_current_Asymm_JJ} (a,d), so that $J_{\Im}$ is sizable and smooth in the entire phase range, as well as the CPR, and we will see that this allows for an easier detection of this current component.

\subsubsection{2G-ZES}

After the G-ZES region, for intermediate asymmetries, e.g. $\gamma=0.5$, the external ZE-EP and the FE-EP appear in the spectrum almost at the same time. Differently, for highly asymmetric JJs, $\gamma=0.8$, the FE-EP manifest in the spectrum \emph{after} the annihilation of the external ZE-EP at $\phi=0$, see the brown thick and dashed curves in the upper panel of Fig.~\ref{fig: 5_Panel_ABS_current_Asymm_JJ}.
In this case a new narrow phase window where no EPs is available where two pairs of G-ZES are present in the spectrum, e.g. \emph{2 G-ZES} region. 
In this limit situation reported in Fig.~\ref{fig: 5_Panel_ABS_current_Asymm_JJ} (b,e), the phase dependence of all the quasi-ABS is encoded in their imaginary part, and thus $J_{\Im}$ is the only supercurrent contribution.

\subsubsection{no ZES (quartet)}

By further increasing $\gamma_{N}$ the system exits the 2 G-ZES region and, at the annihilation point of the FE-EP, reaches a region with no EPs and ZES, i.e. the third brown-shaded region in Fig.~\ref{fig: 5_Panel_ABS_current_Asymm_JJ}, where the imaginary part is still phase-dispersive. This \emph{quartet region} is achieved when one of the spins on the dot is almost decoupled from F, e.g. $\Gamma_{\uparrow}\approx\Gamma_{N}$ and $\Gamma_{\downarrow}\approx0$. This is the case of a strongly polarized ferromagnet~\cite{Grein2009, Bobkova2017, Ouassou2017} or, in the limit of $\gamma_{N}=\Gamma_N$, as the case of a half-metallic ferromagnet~\cite{deGroot1983, Katsnelson2008, Eschrig2008, Eschrig2015, Keizer2006}, see App.~\ref{app: GF_ferromagnetic_lead}.
An example of the ABS spectrum in this case along with the corresponding CPR is provided in Figs.~\ref{fig: 5_Panel_ABS_current_Asymm_JJ} (c,f). 
Interestingly, we notice that here the Andreev levels are found in pairs at the same finite energy but with different broadening everywhere. This is a consequence of the symmetry properties of the system Hamiltonian that we analyze in Sec.~\ref{sec: 4_symmetries}.

\subsection{Experimental measurement of \texorpdfstring{$J_{Im}$}{JIm}}
\label{sec: 3_measurability}

All together these regimes represent situations where $J_{\Im}$ is smooth over the entire phase range and sizable. As we discuss below, the absence of EPs in the spectrum allows a clean detection of such component, especially in the 1 G-ZES regions. 

\subsubsection{Measurement protocol}

\begin{figure*}
    \centering
    \includegraphics[scale=0.207]{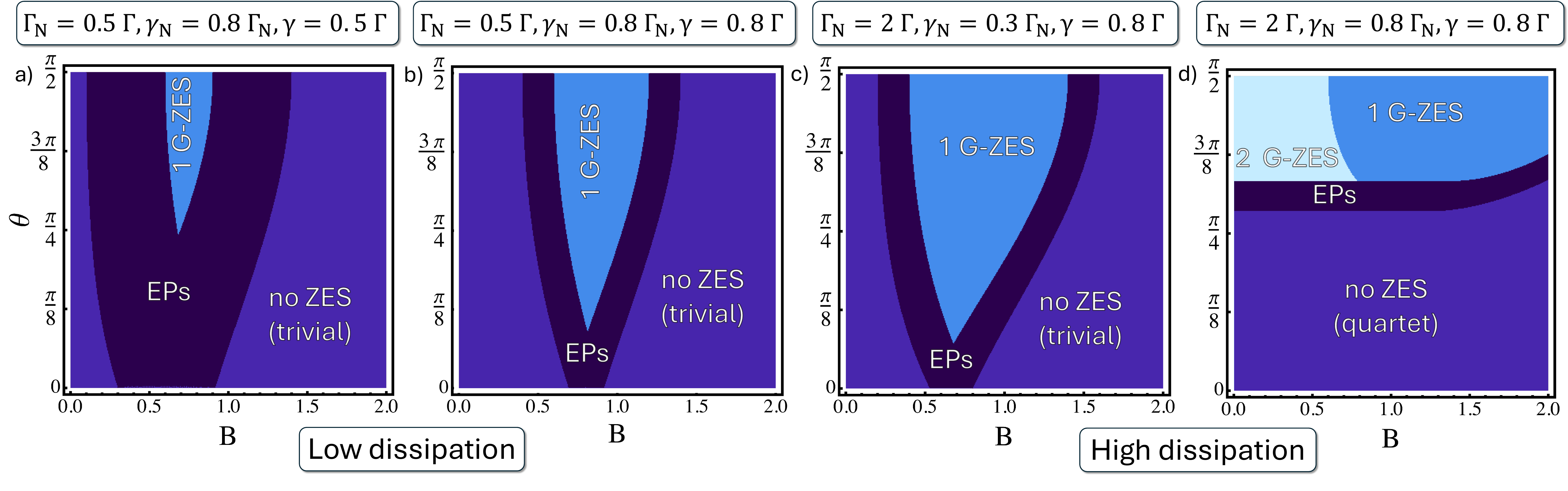}
    \caption{Regime diagrams of the JJ vs. the amplitude and orientation of the field, i.e. $B$ and $\theta$. Regions hosting EPs and G-ZES are reported. In (a-b) we report the case of a JJ with fixed dissipation and spin dissipation imbalance, i.e. $\Gamma_{N}=0.5\Gamma$ and $\gamma_{N}=0.8\Gamma_{N}$, and with different asymmetry parameters, i.e. $\gamma=0.5\Gamma$ and $\gamma=0.8\Gamma$.
    In (c-d) we report the case of a JJ with fixed dissipation and junction asymmetry, i.e. $\Gamma_{N}=2\Gamma$ and $\gamma=0.8\Gamma$, and with different spin dissipation imbalance, i.e. $\gamma_{N}=0.3\Gamma_{N}$ and $\gamma_{N}=0.8\Gamma_{N}$.}
    \label{fig: 6 ZES_Map_total}
\end{figure*}

$J_{\Im}$ can be detected experimentally by the means of combined Josephson current $J(\phi)$ and conductance $dI/dV$ measurements, performed by varying the phase bias $\phi$ as in the scheme of Ref.~\cite{Solow2025}.

Specifically, via $dI/dV(\phi)$ spectroscopy one can easily access at any $\phi$ the real part of energy levels from the peaks of the response. The measurement of the imaginary part is more problematic as the effective width of the $dI/dV(\phi)$ spectroscopic peaks may result from the combination of intrinsic and thermal broadening. Nonetheless, the latter is often negligible at the temperatures relevant for current experiments, as we discuss below. For this reason, we may assume it is a viable measure.
In general, once the spectrum is retrieved from the experiment, the supercurrent formula Eq.~\ref{Jpol_general} can be used to obtain an indirect measurement of both the $J_{\Re}$ and $J_{\Im}$ contribution to the CPR. The result can finally be confronted with the direct measurement of the CPR.

It is not difficult to see that the \emph{2 G-ZES} regime seem to offer a unique chance for probing $J_{\Im}$ since in this case standard CPR measurements would result in a finite supercurrent whereas the spectroscopy would suggest $J_{\Re}=0$ everywhere.
Nonetheless, this regime is achievable only in strongly asymmetric JJs, $\gamma\lesssim 1$, at high values of the average dissipation $\Gamma_{N}$ and high dissipation imbalance between the two spin channels $\gamma_{N}$. The same conditions on the parameters hold for the \emph{quartet} regime. On the contrary, \emph{1 G-ZES} state is achieved also at lower dissipation and junction asymmetry, see the spectrum in Fig.~\ref{fig: 1_EPs_in_ABS_spectrum} (b) that is obtained for $\Gamma_{N}=0.5\Gamma$, $\gamma_{N}=\Gamma_{N}$ and $\gamma=0.5$. 

Transport regimes with low dissipation are in fact desirable for $J_{\Im}$ detection, since for $\Gamma_{N}\lesssim\Gamma$ the system shows at the same time a sizable supercurrent and well-defined Andreev resonances in the $dI/dV(\phi)$, cf. Fig.~\ref{fig: 1_EPs_in_ABS_spectrum}~(b) and Ref.~\cite{Solow2025}.

Further, the \emph{1 G-ZES} configuration represents the most realistic sweet spot for revealing $J_{\Im}$. Here, the pair of ZES contributes only to $J_{\Im}$, while the other pair of eigenvalues at finite real energy only to $J_{\Re}$, having only a constant broadening $\Gamma_{N}$.
Therefore, to determine the existence of a non-trivial $J_{\Im}$, it suffices to determine from the spectroscopy only the contribution from the latter pair and show a discrepancy with the direct CPR measurement. 
We have to mention that the determination of $J_{\Re}$ requires the measurement of the broadening, $\lambda(\phi)$, of the finite energy levels, cf.  Eq.~\ref{Jpol_T0_simp}, which may be subtle due to the aforementioned difficulty. 
However, at dilution temperatures (of order $\sim 10 \,\mathrm{mK}$) the thermal broadening is much smaller than the intrinsic one, due to the coupling to the normal metal~\cite{Beenakker1991, Zhu2001}, so that the width of the Andreev resonances should represent a good estimation of $\lambda(\phi)$. In principle, if the finite energy levels exhibit a phase-dependent broadening, one should measure $\lambda(\phi)$ at each phase bias to calculate $J_{\Re}$.
At resonance, the broadening reduces to $\lambda=\Gamma_{N}$, which not only allows for a single-phase measurement but also provides an alternative route for its determination. Indeed, the ABS inherit the average linewidth of the quantum dot levels when the system is not superconducting and, thus, $\lambda$ might also be retrieved by conductance measurements in the normal state~\cite{Gramich2015, Gramich2017}.

As a proof of the suitability of the \emph{1 G-ZES} regime for the detection of $J_{\Im}$, we recall the results shown in Fig.~\ref{fig: 1_EPs_in_ABS_spectrum}(b). 
In this case, the system can host a $J_{\Im}$ contribution of up to $\sim 30\%$ of the maximum current, with $J_{\mathrm{tot}}(\phi=\pi/2)$ being about $50\%$ larger than $J_{\Re}(\phi=\pi/2)$. 
Such a large discrepancy should be experimentally detectable assuming an uncertainty on $J_{\mathrm{tot}}$ of order $\sim 5$--$10\%$ of $J_{\mathrm{tot}}(\phi=\pi/2)$~\cite{DellaRocca2007, Basset2014, Delagrange2016, Ahmad2022}, and a comparable uncertainty on $J_{\Re}$ extracted from $dI/dV(\phi)$ measurements~\cite{Nichele2020}.

We remark that the aforementioned desired conditions for $J_{\Im}$ detection, realized in the \emph{1 G-ZES} regions, Fig.~\ref{fig: 1_EPs_in_ABS_spectrum}~(b), are in principle achieved also in the EPs regime within the phase window featuring the ZES line, Fig.~\ref{fig: 1_EPs_in_ABS_spectrum}~(a). 

However, this regime is found less suitable for probing $J_{\Im}$, especially when the two ZE-EPs are close to each other. Indeed, a reasonable width for the zero-energy line, where only the two finite-energy levels contribute to $J_{\Re}$, is desirable so as to ease the estimation of $J_{\Re}$. Unfortunately, since the position of the EPs depends strongly on the system parameters, it is hard to engineer the sought Andreev spectrum.

In this respect, the \emph{1 G-ZES} regions represent a controlled environment where such issues are circumvented. Here, the Andreev spectra feature no EPs and importantly small perturbations in the system parameters do not affect the spectral structure.

\subsubsection{Experimentally accessible regimes}

In order to investigate if the state of the JJ can be easily tuned between the different regimes, e.g. hosting EPs as well as G-ZES, we explored the ABS spectral configurations by varying the orientation and amplitude of the external magnetic field, $B$ and $\theta$. 
This investigation is motivated by the fact that the position of the EPs depend also on these parameters, Eqs.~\ref{EPs_position}, that can be more easily tunable in a possible experiment.
We report the results in the \emph{regime diagrams} in Figs.~\ref{fig: 6 ZES_Map_total}
where we indicate the regions hosting EPs and G-ZES and those with no ZES in the spectrum.

The \emph{no ZES} regions comes in two kinds. The first one, named \emph{trivial}, hosts spectral configurations with constant imaginary part of occasionally hosting FE-EPs. We note that any gapped Hermitian spectra will fall into this type of region when the dissipation is turned on. For this reason they are realized either at small or large magnetic fields.
The second kind is related to the aforementioned \emph{quartet} structure.

These results confirm that the \emph{1 G-ZES} state could be easily accessed by controlling the external magnetic field both at low and high dissipation, Figs.~\ref{fig: 6 ZES_Map_total} (a-d). The \emph{1 G-ZES} region is sizably enhanced for the highly asymmetric JJs, Fig.~\ref{fig: 6 ZES_Map_total} (b), and is we verified it is often larger at high $\Gamma_{N}$, e.g. Fig.~\ref{fig: 6 ZES_Map_total} (c). Instead, the \emph{2 G-ZES} and \textit{quartet} regions are visible only at high dissipation, imbalance and asymmetry, Fig.~\ref{fig: 6 ZES_Map_total} (d).
We note that a tiny \emph{trivial} \emph{no ZES} spot also appear in Fig.~\ref{fig: 6 ZES_Map_total} (d), but is not shown for the sake of clarity.

As a final remark, we got evidence that an orthogonal configuration with $\theta\sim\pi/2$ is in general favorable to realize \emph{1 G-ZES} phases.

\section{ZES intervals as symmetry broken phases}
\label{sec: 4_symmetries}

We have shown in Sec.~\ref{sec: 3_Asymmetric_JJ_in_Res} that in the resonant regime imaginary parts of the quasi-ABS levels become phase dispersive in presence of ZE- or FE-EPs. Indeed, EPs lie at the boundaries between regions where the states have different real energy and the same \emph{constant} imaginary part ($\Gamma_{N}$) on one side and regions where the levels are paired together at the same real energy with a different phase-dependent broadening on the other, see Fig.~\ref{fig: 4_Prototype_spectrum}. 
As we mentioned in Sec.~\ref{sec: 3_analytic_ABS} this phase-dispersive behavior of the levels imaginary part can be read as a consequence of the effective Hamiltonian symmetries.
For what concerns the ZE-EPs, when a pair of particle hole symmetric levels $(z=\varepsilon-i\lambda,-z^{*}=-\varepsilon-i\lambda)$ coalesces on the real axis, $\Re\{z\}=0$, a transition between different PHS$^{\dagger}$ phases occurs within the corresponding two-level subspace and, after the EP, the pair disassembles into two unpaired levels (with phase-dispersive imaginary parts), $(i\lambda_{+}, i\lambda_{-})$, see Appendix~\ref{app: symmetries}.
This does not apply to the FE-EPs, where the eigenvalues (and the eigenvectors) involved are not related by any symmetry of $\check{H}_{eff}$ (listed in Tab.~\ref{tab: Hsyms} of Appendix~\ref{app: symmetries}).
 
Still we note that the phase dispersion in the imaginary part of quasi-ABS sets on again exactly after these spectral points, Fig.~\ref{fig: 4_Prototype_spectrum}.

The spectral properties of 
$\check{H}_{eff}$ can be understood at full when considering a shifted Hamiltonian where the channel-averaged decaying background is removed,
\begin{equation}            
\label{reduced_Hermitian_H}
\check{H}'(\phi)=\check{H}_{eff}(\phi)+i\Gamma_{N}\check{1}\,,
\end{equation}
as done typically in optics when dealing with NH passive systems~\cite{Pen14, Bittner2012, Ornigotti_2014, Ozdemir2019, Savarese2025}.

\begin{figure}[h]
	\centering
	\includegraphics[scale=0.72]{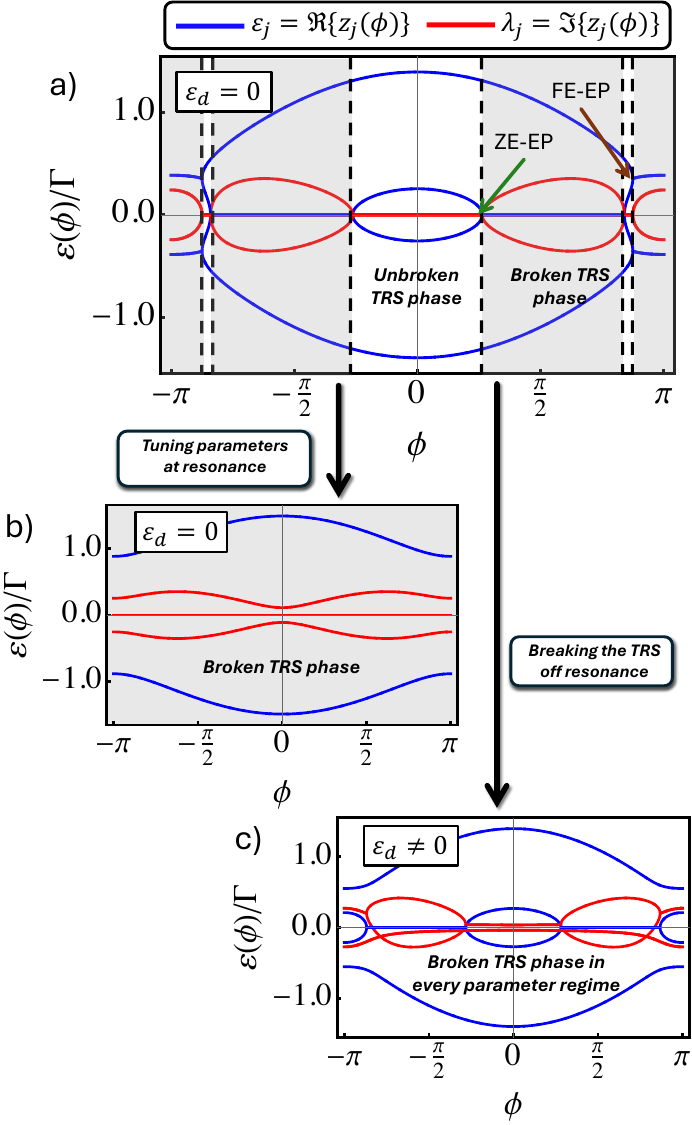}
	\caption{Spectrum of the shifted Hamiltonian $H'$ for a JJ in resonance, $\varepsilon_d=0$, hosting EPs (a) and a pair of G-ZES (b), and out of resonance, $\varepsilon_d=0.1$, hosting EPs (c).
    In (a) and (b), $H'$ features TRS and \emph{broken} and \emph{unbroken} regions of the spectrum can emerge. Broken regions are shaded in gray and their borders, characterized with EPs, are delimited by black dashed lines.  The other system parameters are $\Gamma=1$, $\theta=\pi/4$, $\gamma_N=0.5$,  $\gamma=0.3$, $B=0.5$ (a-b), and  $\gamma=0.5$, $B=0.6$ (c).}
	\label{fig: 8_ABS_spectrum_epsdne0_epsdequal0}
\end{figure}

This Hamiltonian is more symmetric than $\check{H}_{eff}$, despite the only difference is a constant shift of band spectrum, see the end of Tab.~\ref{tab: Hsyms}. 
Crucially an additional (anti-unitary) Time-Reversal-like Symmetry (TRS) is found for $\check{H}'$ in the resonant tunneling regime, i.e. $\varepsilon_{d}=0$, that can explain both the presence of FE-EPs in the spectrum~\cite{Flensberg2021, Solow2025} and the peculiar phase dependence of the levels imaginary part.
We remark that, since the NH symmetry classes are defined up to a shift and a multiplication of the Hamiltonian by complex constants, here $\check{H}'$ represents the right object for a full characterization of the symmetries~\cite{Kawabata2019}. The additional TRS symmetry ensures that for each eigenvalue $z'=\varepsilon-i\lambda'$ another state with $z'^*=\varepsilon+i\lambda'$ is present. In the contemporary presence of PHS$^{\dagger}$ and TRS, the eigenvalues of $\check{H}'$ exist either in pairs of particle-hole symmetric levels with unrelated real parts but vanishing imaginary parts, of the kind $(\varepsilon,-\varepsilon)$, or in complex conjugated pairs, with the EPs being the transition points between the two phases. More specifically, we note that in the intervals between two ZE-EPs, central white region and adjacent gray regions in Fig.~\ref{fig: 8_ABS_spectrum_epsdne0_epsdequal0} (a), the levels appear respectively in two pairs $(\varepsilon,-\varepsilon)$ and $(\lambda',-\lambda')$, while after a FE-EP (external gray regions) the spectrum is made of quartets $(z',z'^*,-z',-z'^*)$. Further details on the effect of the TRS on $H'$ on the eigenvalues of $H_{eff}$ are analyzed in Appendix~\ref{app: symmetries}.

Analogously to PHS$^{\dagger}$, the existence of two TRS phases is due to the anti-unitarity of its associated operator.
Moreover, being characterized by either a totally real spectrum or one featuring complex conjugated pairs, such symmetry is an analog of $\mathcal{PT}-$symmetry and we shall refer to these phases respectively as \emph{unbroken} and \emph{broken}~\cite{Bender07, Bender:1998, Gong_2020,Kawabata2019, Kawabata2018}, see Fig.~\ref{fig: 8_ABS_spectrum_epsdne0_epsdequal0}(a).

Crucially, such a combination of symmetries of $\check{H}'(\phi)$ allows us to see a non-trivial property of $\check{H}_{eff}(\phi)$. The spectrum of $\check{H}'(\phi)$ is symmetric with respect to the 0-axis, see Fig.~\ref{fig: 8_ABS_spectrum_epsdne0_epsdequal0} (a), i.e. it is symmetric with respect to both real and imaginary axes.
In the unbroken regions of TRS, the imaginary part of $\check{H}_{eff}$ eigenvalues cannot depend on the phase $\phi$ and be responsible of supercurrent.

Therefore, from a practical point of view, what we have done in Sec.~\ref{sec: 3_spectral_regimes_GZES} is to tune the system parameters so as to extend the broken regions of the TRS to all phase biases. \emph{1 G-ZES} regions can be simply seen as spectral configurations where the TRS is spontaneously broken everywhere and thus the $J_{\Im}$ contribution of ZES is always present and smooth, see Fig.~\ref{fig: 8_ABS_spectrum_epsdne0_epsdequal0} (b).
We observe that this quartet structure also explains the peculiar \emph{quartet} regime in Fig.~\ref{fig: 5_Panel_ABS_current_Asymm_JJ}. This is indeed a phase where the levels are paired together at the same finite energy in the whole phase range. This regime can be obtained when the FE-EPs annihilate at $\phi=0$ instead of $\phi=\pi$.

However, TRS on $\check{H}'$ is naturally lost when the dot is off-resonance, $\varepsilon_d\neq0$. In Fig.~\ref{fig: 8_ABS_spectrum_epsdne0_epsdequal0} (c) we show how the spectrum of $\check H'$ changes when the system exits the resonant tunneling regime testifying that the imaginary part of the levels, $\lambda(\phi)$, becomes dispersive everywhere.

In this framework, the non-resonant tunneling regime promises an even more enhanced $J_{\Im}$ contribution to the junction CPR. In this limit, all the quasi-ABS contribute to $J_{\Im}$, at the cost of measuring the imaginary part of the levels at each phase to detect $J_{\Im}$. 
We explore the parameters regimes in this more general situation in App.~\ref{sec: 5_Off_res_JJ} in order to find out whether optimal configurations for the experiments, such as the \emph{1 G-ZES} region, can be still achieved.

\section{Summary and Conclusions}
\label{sec: 6_Conclusions}

\par

In conclusion, we investigated the transport properties of NH JJs with a single-level quantum dot barrier coupled to a ferromagnetic reservoir under an external magnetic field. The object of our analysis was the additional supercurrent contribution proportional to the 
phase derivative of the imaginary part of the Andreev levels.
Using the NH current formula that we derived in Ref.~\cite{Capecelatro2025}, we identified regimes where this NH current contribution($J_{\Im}$) can be easily resolved and we outline a possible experimental protocol for its detection.
We analyzed both the resonant and non-resonant tunneling regimes.

In resonant tunneling, we found analytic expressions of both the quasi-ABS and the EPs positions thus enabling full manipulation of the Andreev spectrum by tuning the junction asymmetry, the spin-dependent dissipation, and the magnetic field.
Andreev spectra hosting one pair of globally extended ZES emerge as optimal for probing $J_{\Im}$. Here the ZES contribute solely to $J_{\Im}$ while the finite-energy states contribute only through their real part, $J_{\Re}$. 
Any discrepancy between the $J_{\Re}$ computed from the $dI/dV(\phi)$ spectroscopy as in Refs.~\cite{Capecelatro2025, Beenakker2024} and the measured CPR would thus verify the presence of the $J_{\Im}$ contribution.
Our findings demonstrated that exotic NH Andreev spectra can be obtained in a remarkably simple s-wave device, hinting that also more complex but versatile junctions including spin-orbit coupling~\cite{Zazunov2009, Campagnano_2015, Minutillo2018, Guarcello2024, Maiellaro2024} and interactions~\cite{Rozhkov1999, Cuevas2001, Flensberg2021} may feature similar spectra. 

At a deeper level, in our work we showed that the phase dispersion in the imaginary part of the levels, together with $J_{\Im}$, arises in the broken phases of the TRS in the shifted Hamiltonian $H'$, where the average loss term is subtracted.
We therefore highlighted the connection between the symmetries of the Hamiltonian and the physical supercurrent.
This remarkable phenomenon could be investigated in other transport observables, such as the thermal current and the current noise, and beyond the equilibrium regime investigated here and may guide the design of solid-state systems where NH effects are exploited at full. We believe such considerations represent a promising direction for further studies. 
We noted that, despite the constant term $-i \Gamma_N$ being essential for computing physical observables, it leaves the ABS unaffected and the true \emph{source of non-Hermiticity} is the dissipation imbalance ($\gamma_N$). Therefore, being able to control all sources of decoherence separately is essential for reproducing in NH Josephson devices those peculiar NH features that are engineered in optical systems, e.g. through balanced gains and losses \cite{Ramezani2010,El-Ganainy2018, Chang2014}.

In the non-resonant regime the TRS is explicitly broken and all quasi-ABS acquire a phase-dispersive broadening so that the visibility of $J_{\Im}$ generally improves, see Appendix~\ref{sec: 5_Off_res_JJ}. Off-resonance,  both zero-energy and finite-energy states contribute to $J_{\Im}$, thus making the extraction of $J_{\Im}$ from the CPR less straightforward.
In fact, the detection protocol in this regime involves the measurement of the levels broadening at all phase-biases, thus being more demanding with respect to the resonant case.
However, this regime does not require fine tuning of the gate on the dot.
As a further step toward a more realistic regime of parameters, we envision the extension to finite gap junctions, beyond the infinite-gap limit assumed in the present analysis. This physics can be still explored by means of our general current formula~\cite{Capecelatro2025}, Eq.~\ref{Jpol_general}.

Our results may also be relevant for photonic analogues of Josephson junctions~\cite{Lagoudakis2010, Abbarchi2013, Voronova2025}. Driven–dissipative photonic systems, such as exciton–polaritons condensates, which already exhibit Josephson oscillations, have been predicted to host synthetic Andreev bands and Andreev-like states~\cite{Septembre2021, Septembre2023, Septembre2023_MultiTerminal}. 
These platforms are intrinsically non-Hermitian due to gains and losses~\cite{Pickup2020}, that could be engineered to investigate non-Hermitian Andreev physics in a controlled environment, possibly including the non-Hermitian supercurrent contribution discussed in this work.

Overall, our work showed that genuine NH effects in transport observables can be detected and controlled in simple s-wave JJs, without requiring the presence of EPs in the spectrum. This significantly broadens the experimental accessibility of NH physics, by moving beyond the idea, assumed in previous proposals, that EPs constitute the sole mechanism to access NH transport signatures.
\begin{acknowledgments}
This work was supported by PNRR MUR project~PE0000023 - NQSTI (TOPQIN - CUP J13C22000680006), by the European Union's Horizon 2020 research and innovation programme under Grant Agreement No~101017733, by the MUR project~CN\_00000013-ICSC, by the  QuantERA II Programme STAQS project that has received funding from the European Union's Horizon 2020 research and innovation programme under Grant Agreement No~101017733, and by PRIN MUR Project TANQU 2022FLSPAJ. 
This work was partially supported by Horizon Europe EIC Pathfinder under the grant IQARO number 101115190.
R. Capecelatro and R. Citro acknowledge R. Aguado, C. Guarcello, F. Romeo and M. Trama for fruitful discussions. 
\end{acknowledgments}
\appendix
\section{Off-resonance regime and stability of the Global ZES}
\label{sec: 5_Off_res_JJ} 
As we discussed in Sec.~\ref{sec: 4_symmetries}, in the off-resonance regime the TRS-like  symmetry in the shifted Hamiltonian, $\check{H}'$, is explicitly broken and the imaginary part of the levels becomes phase-dispersive even in the absence of EPs or ZES. 
When numerically sampling the $(B,\theta)$ space, we find regions where the ABS spectrum host neither EPs nor ZES but the system is characterized by a smooth and sizable $J_{\Im}$ contribution, see Fig.~\ref{fig: 8_ABS_CPR_epsdneq0} (a, b).
By properly tuning $(B, \theta)$ the \emph{1 G-ZES} regime is retrieved also in this case, see Fig.~\ref{fig: 8_ABS_CPR_epsdneq0} (c, d).

\begin{figure}[h]
	\centering
	\includegraphics[scale=0.23]{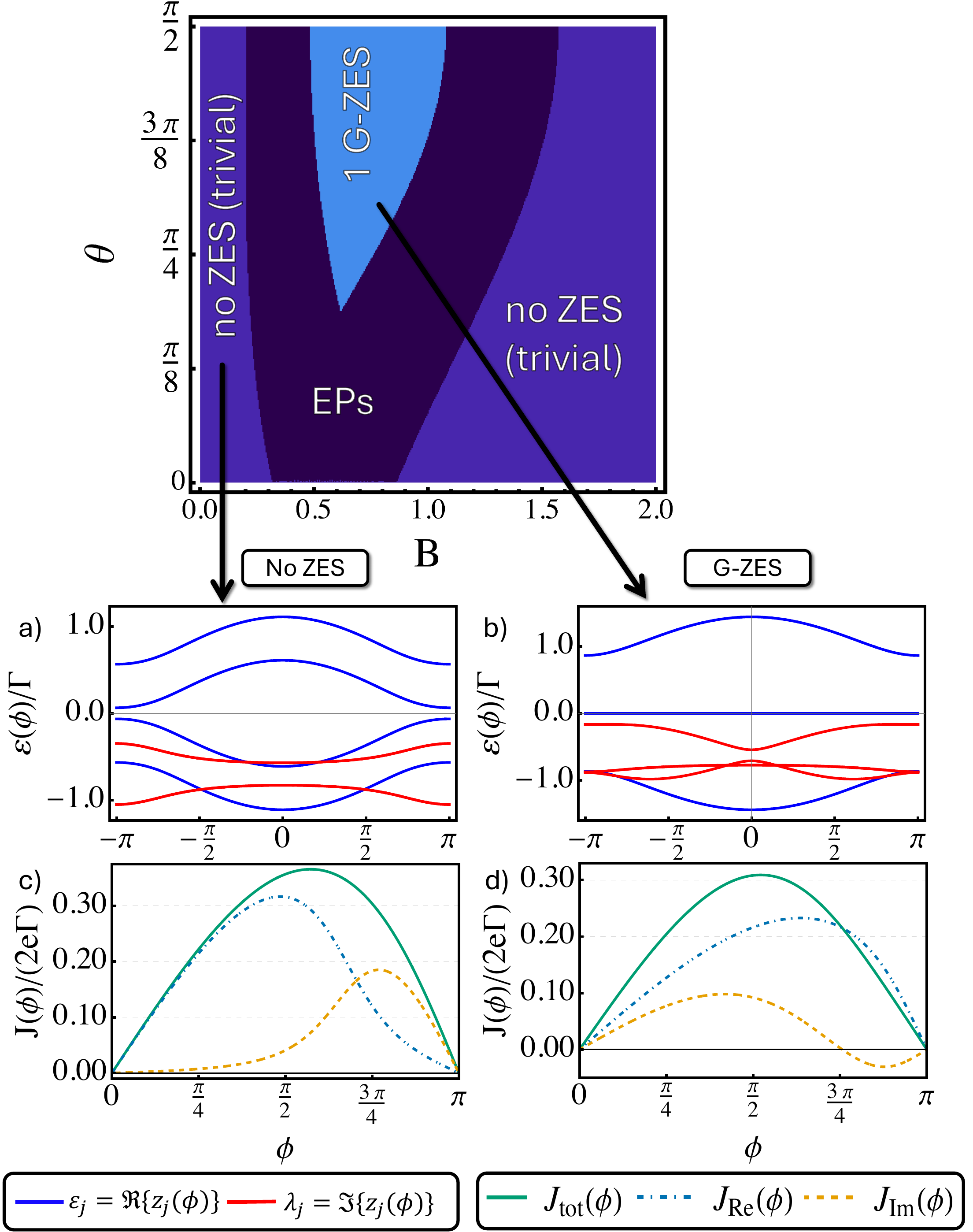}
	\caption{Regime diagram of a JJ in the off-resonance limit with respect to the magnetic field strength and orientation $(B,\theta)$ (upper panel) and typical spectra (a-b) and CPRs (c-d). The system parameters in the upper panel are $\varepsilon_{d}=0.2$, $\Gamma=1$, $\gamma=0.3$, $\Gamma_{N}=0.7$ and $\gamma_N=0.8\Gamma_{N}$. Two situations are reported keeping the notation of Fig.~\ref{fig: 5_Panel_ABS_current_Asymm_JJ}: no ZES nor EPs are present (a, c); the spectrum features one pair of G-ZES (b, d). The parameters are respectively $(B=0.1,\theta=0.2)$, for (a, c), and $(B=0.5,\theta=1.3)$, for (b, d).}
    \label{fig: 8_ABS_CPR_epsdneq0}
\end{figure}

We note that while the \emph{2 G-ZES} regime, Fig.~\ref{fig: 5_Panel_ABS_current_Asymm_JJ} (b), 
can be still realized very close to the resonance condition $\varepsilon_{d}\ll\Gamma$, the quartet regime, Fig.~\ref{fig: 5_Panel_ABS_current_Asymm_JJ} (c), cannot be achieved in the absence of TRS on $H'$, see Appendix~\ref{app: symmetries}.

Our numerics proves that in the off-resonance regime the phase-dispersive behavior of the quasi-ABS imaginary part is in general more pronounced. Moreover, the regions of the parameter space where $J_{\Im}$ might be efficiently highlighted in the experiments, i.e. where it is smooth and comparable to $J_{\Re}$, are more extended.

A convenient dimensionless parameter to evaluate the fraction of the contribution to the CPR of the imaginary-part dispersion is
\begin{eqnarray}
  \label{eq:eta_def}
  \eta
  &=& \frac{I_{\Im}}
         {I_{\Re} + I_{\Im}} \quad\quad(0\leq \eta \leq 1),\\
    I_{\Re/\Im} &=&\int_{0}^{2\pi} d\phi\,\big|J_{\Re/\Im}(\phi)\big|\nonumber
\end{eqnarray}
where $I_{\Re/\Im}$ is an L1-norm of the current.
The choice of such norm owes to its stability at EPs, and the fact that it weights equally all phases, as opposed to a "max" measure.

\begin{figure}[h]
	\centering
	\includegraphics[scale=0.45]{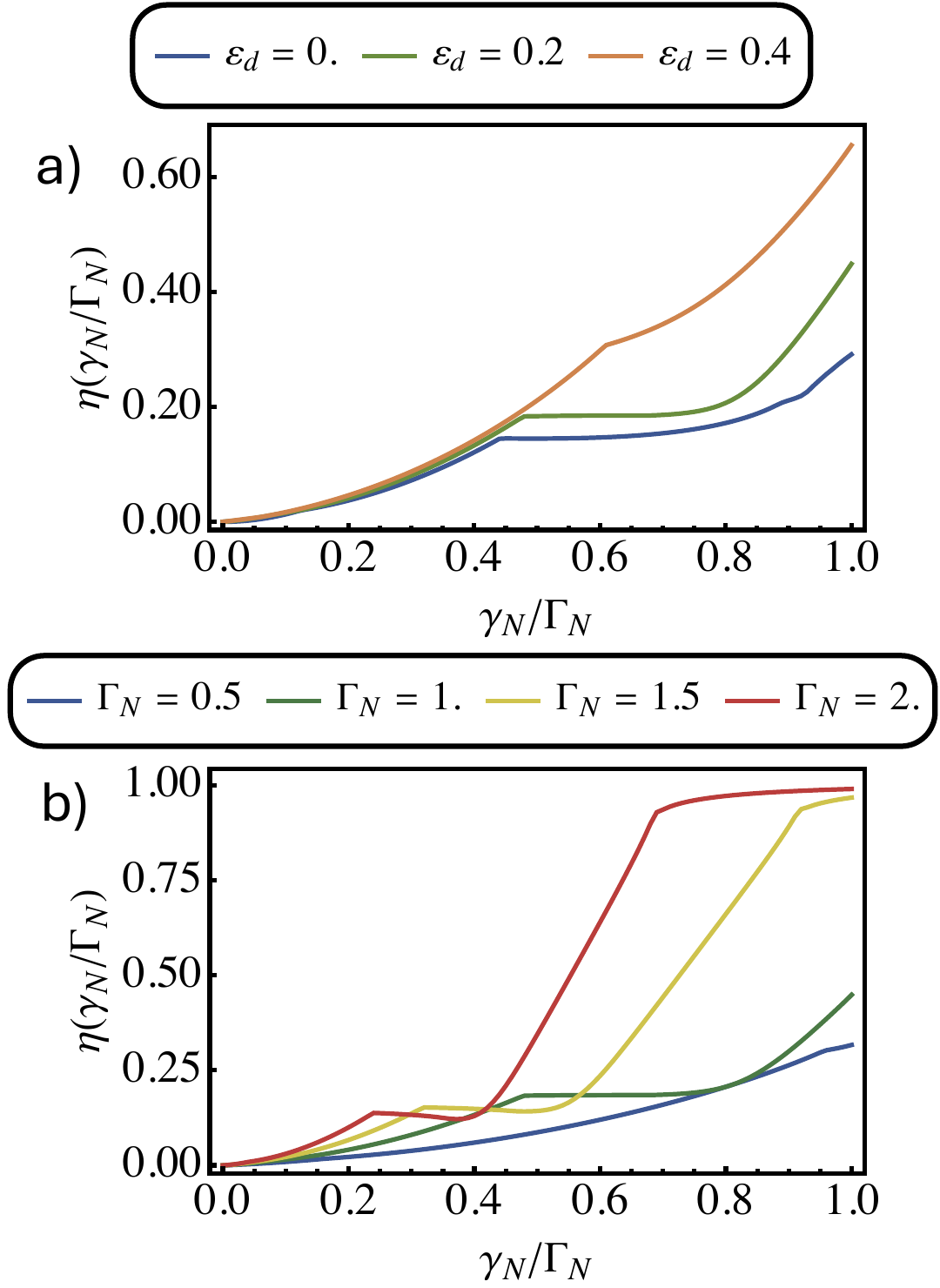}
	\caption{The fraction $\eta$, cf. Eq.~\ref{eq:eta_def}, as a function of the relative spin dissipation imbalance $\gamma_{N}/\Gamma_{N}$ for different values of the dot energy $\varepsilon_{d}$ (upper panel, (a)) and different values of the average coupling to the F lead $\Gamma_{N}$ (lower panel, (b))
    The system parameters are $\Gamma=1$, $\gamma=0.5$, $\Gamma_{N}=1$, $B=0.6$ and $\theta=1$ (a) and $\Gamma=1$, $\gamma=0.5$, $\varepsilon_{d}=0.2$, $B=0.6$ and $\theta=1$ (b).}
    \label{fig: 8_ABS_CPR_epsdneq0}
\end{figure}

In Fig.~\ref{fig: 8_ABS_CPR_epsdneq0} we show $\eta$ as a function of the dissipation imbalance between the two spins $\gamma_{N}/\Gamma_{N}$ for different values of the dot energy $\varepsilon_{d}$ (a) and different values of the average dissipation $\Gamma_{N}$ (b).
We observe that in general the higher $\gamma_{N}/\Gamma_{N}$ the higher $\eta$. This result reflects
the fact that $\gamma_{N}$ measures the \emph{degree of the eigenvectors non-Hermiticity} of the system and that for $\gamma_{N}=0$ there is no dispersion in the imaginary part of the quasi-ABS. Indeed, in the latter case non-Hermiticity manifests only as an imaginary constant shift of the spectrum while the eigenvectors structure stays the same of the Hermitian case.
Further, in general having a non-zero voltage bias, tuning up $\varepsilon_{d}$, enhances $\eta$, however this effect is sizable only for large $\gamma_{N}$, e.g. $\gamma_{N}>0.5\Gamma_{N}$.
This can be explained by the fact that the more the dot is tuned out of resonance the lower will be the transparency of the channels and the flatter will become the real part of the levels, as usual for closed junction, resulting in  $J_{\Re}\lesssim J_{\Im}$.

We notice that the $\eta$ curves may feature a kink where the system enters the \emph{1 G-ZES} regime and they develop a plateau for the whole length in $\gamma_{N}$ of this region.
Differently, increasing the average dissipation $\Gamma_{N}$ raises the ratio $\eta$ only at high values of $\gamma_{N}$ whereas at smaller values its effect is strongly dependent on the system. Moreover, for large $\Gamma_{N}$ the $\eta$ curves show a second plateau at $\eta\lesssim1$. These regions correspond to situations where again one pair of G-ZES is present in the spectrum but the other two quasi-ABS at finite energy have an almost flat real part so that 
$J_{\Re}\rightarrow0$. Then, we note that the higher $\Gamma_{N}$ the earlier the appearance of G-ZES plateaus in the $\eta$ curves.
As a general remark on the off-resonance regime, we point out that since all the quasi-ABS have  phase-dispersive imaginary part we would need to sample the levels broadening at all phase biases to have a correct estimation of $J_{\Re}$ and hence probe $J_{\Im}$. Finally, we observe that whereas in the resonant tunneling the detection protocol might be easier, in the non-resonant tunneling the effect of $J_{\Im}$ on the CPR is larger, thus, making the choice of the regime for the measurements dependent on the experimental setup.

\section{Green's function of the ferromagnetic lead}
\label{app: GF_ferromagnetic_lead}
In this Appendix, we report the Matsubara GF of the F lead. We model it as a normal 1D semi-infinite chain in tight-binding formalism, following Ref.~\cite{Ferry2009:book}, with a finite Zeeman splitting, $h=|\vec{M}_{F}|$, due to the F lead magnetization. For the sake of simplicity we choose $z$ as the spin quantization axis of the system.
The tight-binding Hamiltonian for the ferromagnet reads
\begin{eqnarray}
H_{\rm F} &=&
\sum_{m,\sigma=\uparrow,\downarrow}\left[
\mu_N  c_{m,\sigma}^\dagger c_{m,\sigma}+\left(\tau_N\, c_{m+1,\sigma}^\dagger c_{m,\sigma}
+ {\rm H.c.}\right)\right]+
\nonumber\\
&&
+ \sum_m \left[
h \left( c_{m,\uparrow}^\dagger c_{m,\uparrow}- c_{m,\downarrow}^\dagger c_{m,\downarrow} \right)
\right] =\\
&=&\sum_{k,\sigma=\uparrow,\downarrow} \left[
\left(\mu_N - \varepsilon_{k}\right)
c_{k,\sigma}^{\dagger}c_{k,\sigma}\right]+\sum_{k}h\left(
c_{k,\uparrow}^{\dagger}c_{k,\uparrow}
- c_{k,\downarrow}^{\dagger}c_{k,\downarrow}
\right),\nonumber
\end{eqnarray}

with $\mu_{N}$ being the normal metal chemical potential and $\varepsilon_{k}=-2 \tau_{N} \cos{\left(k a_{N}\right)}$ being the dispersion law in momentum space. $\tau_{N}$ and $a_{N}$ are the hopping parameter and lattice constant of the chain, respectively.

The influence of the F lead can be condensed in the surface GF of the semi-infinite chain, that is the GF at its last site, which is connected to the QD.
This GF can be computed via recursive Green's function methods~\cite{Capecelatro2025, Ferry2009:book, Datta_1995, Furusaki1994, Asano2001, Minutillo2021} and can be simply expressed by a diagonal matrix in spin space
\begin{equation}
\hat{G}_{F} =
\begin{pmatrix}
G_{\uparrow}& 0 \\
0 & G_{\downarrow}
\end{pmatrix},
\end{equation}
whose elements are 
\begin{eqnarray}
G_{\uparrow/\downarrow}\left(i\omega_{n}\right)&=&\frac{\left(i\omega_{n}-(\mu_{N}\pm h)\right)}{2\tau^{2}_{N}}\\
&&-i\frac{\mathrm{sign}(\omega_{n})}{\tau_N}\sqrt{1-\frac{\left(i\omega_{n}-(\mu_{N}\pm h)\right)^2}{4\tau_{N}^2}}\,.\nonumber
\end{eqnarray}
In the broad-band limit $\tau_{N}$ is much larger then any other energy scale of the system a part from the F lead magnetization $\vec{M}_{F}$. Here, the Zeeman term in F, $h$, is still sizable to induce a different imaginary term for the two spins, i.e.
$
G_{\uparrow/\downarrow}\left(i\omega_{n}\right)=-(i/\tau_N)\mathrm{sign}(\omega_{n})\sqrt{1-\left(\mu_{N}\pm h\right)^2/4\tau_{N}^2}$.

The self-energy term accounting for the interactions between the dot and the F lead is formally written as
\begin{equation}
    \hat{\Sigma}_{F}(i\omega_{n})=\hat{H}_{T_{F}}^{\dagger}\hat{G}_{F}\hat{H}_{T_{F}}
\end{equation}
where $\hat{H}_{T_{F}}=t_{F}\hat{\sigma}_{0}$ and $t_{F}$ is the hopping amplitude between the dot and the chain.
Its explicit matrix form in the broad-band limit therefore reads:
\begin{eqnarray}
    &\hat{\Sigma}_{F}&(i\omega_{n})=
\begin{pmatrix}
-i \Gamma_{\uparrow}\mathrm{sign}\left(\omega_{n}\right) & 0 \\
0 & -i\Gamma_{\downarrow}\mathrm{sign}\left(\omega_{n}\right) 
\end{pmatrix}\nonumber\\
&&\mathrm{with}\;\; \Gamma_{\uparrow/\downarrow}=\frac{t_{F}^2}{\tau_N} \sqrt{1-\left(\mu_{N}\pm h\right)^2/4\tau_N^2}\,.
\end{eqnarray}

We note that, the retarded form of this self-energy component, $\hat{\Sigma}_{F}^{R}$, can be rewritten in terms of the spin-resolved densities of states (DOS), $\rho_{\uparrow/\downarrow}$.
Indeed, by definition we have 
\begin{eqnarray}
\rho_{\uparrow/\downarrow}&=&-\frac{1}{\pi}\Im\left[G^{R}_{\uparrow\downarrow}(\omega)\right]=\\
&=& \,\frac{1}{\tau_{N}\pi}
\sqrt{1 - \left(\frac{\omega - \mu_{N}\mp h}{2 \tau_{N}}\right)^{2}}\Theta\!\left(1 - \left|\frac{\omega - \mu_{N}\mp h}{2 \tau_{N}}\right|\right)\nonumber,\\
&\overset{\tau_N\gg\omega}{=}&\frac{1}{\tau_{N}\pi}
\sqrt{1 - \left(\frac{ \mu_{N}\pm h}{2 \tau_{N}}\right)^{2}}\nonumber
\end{eqnarray}
so that 
\begin{eqnarray}
    \hat{\Sigma}_{F}=
\begin{pmatrix}
-i t_{F}^2\rho_{\uparrow} & 0 \\
0 & -it_{F}^2\rho_{\downarrow} 
\end{pmatrix}\; \mathrm{with}\;\;\Gamma_{\uparrow/\downarrow}=\pi t_{F}^2\rho_{\uparrow/\downarrow}\,.
\end{eqnarray}
Hence, the dissipation imbalance, $\gamma_{N}=(\Gamma_{\uparrow}-\Gamma_{\downarrow})/2$, between the two spin-channels arises due to the different occupation in the two spin bands of the ferromagnet.
In this framework, having $\gamma_{N}\lesssim1$ would correspond to have a strongly polarized ferromagnet~\cite{Grein2009, Bobkova2017, Ouassou2017} and in the limit of $\gamma_{N}=1$ a ferromagnetic half-metal \cite{deGroot1983, Katsnelson2008, Eschrig2008, Eschrig2015, Keizer2006}, for which there are no states available for one spin band and the other channel is the only available for the transport. 

\section{Green's function and effective Hamiltonian in the finite \texorpdfstring{$\Delta$}{Delta} case} 
\label{app: GF_finite_Delta_case}
In this appendix we recall the expression for the GF of the QD coupled to the S and F leads in the finite $\Delta$ case and show how to retrieve the effective Hamiltonian in Eq.~\ref{Heff_NH} in the infinite $\Delta$ limit. 

The influence of the S leads and the F bath on the QD can be included by the means of self-energy terms in Matsubara GF formalism, $\Sigma_{S}\left(\omega_{n}\right)=\Sigma_{L}+\Sigma_{R}$ and $\Sigma_{F}(\omega_{n})$.
The dot GF is then $G_{d}\left(\omega_{n}\right)=\left(i\omega_{n}- H_{d}-\Sigma_{S}(\omega_{n}) - \Sigma_{F}(\omega_{n})\right)^{-1}$.

The S leads are described by a flat conduction band with a constant density of state $\rho_{0}$ at the Fermi level~\cite{Meng2009_PRB, Zazunov2009,JonckhereeMartin2009,Benjamin2007,Capecelatro2023}.
By denoting with $\Gamma_{L,R}$ the hybridization parameters between the dot and the left/right S lead, and using $\Gamma_{L}=(\Gamma+\gamma)/2$ and $\Gamma_{R}=(\Gamma-\gamma)/2$, we have 
\begin{eqnarray}
\label{S_leads_self_energy_Delta_finite}
    \check \Sigma_{S}&=&\frac{
     i\omega_{n}\Gamma}{\sqrt{\Delta^2-(i\omega_{n})^2}}\check{1}+\\&&\frac{\Delta}{\sqrt{\Delta^2-(i\omega_{n})^2}} \left[\Gamma\cos{\left(\frac{\phi}{2}\right)}\hat{\tau}_{x}\otimes \hat{1}+\gamma\sin{\left(\frac{\phi}{2}\right)}\hat{\tau}_{y}\otimes \hat{1}\right]\,.\nonumber
\end{eqnarray}
This in the large gap limit, $\Delta\gg\Gamma_{L,R}$, becomes Eq.~\ref{S_leads_self_energy}.

The self-energy term for the ferromagnetic metal reservoir, F, in the broadband approximation is the same of Eq.~\ref{F_lead_self_energy_broadband}.

When using the parametrization $\Gamma_{\uparrow}=\Gamma_{N}+\gamma_{N}$ and $\Gamma_{\downarrow}=\Gamma_{N}-\gamma_{N}$, as in sec.~\ref{sec: 2_Model_GF}, the retarded dot GF in Nambu$\otimes$spin space reads
\begin{eqnarray}
\label{Gmatsu}
\check{G}_{d}(z)& =  \Bigg[& z\left(1+\frac{\Gamma}{\sqrt{\Delta^2-z^2}}\right)\check{1} -\nonumber\\
&&\varepsilon_{d}\hat{\tau}_{z}\otimes\hat{1}-
B_{z}\hat{1}\otimes\hat{\sigma}_{z}-B_{x}\hat{1}\otimes\hat{\sigma}_{x}+\nonumber\\ 
&&i\Gamma_{N}\check{1}+i\gamma_{N}\hat{\tau}_{z}\otimes\hat{\sigma}_{z}-\frac{\Delta}{\sqrt{\Delta^2-z^2}}\times \\
&&\left(\Gamma\cos{\left(\frac{\phi}{2}\right)}\hat{\tau}_{x}\otimes \hat{1}+\gamma\sin{\left(\frac{\phi}{2}\right)}\hat{\tau}_{y}\otimes \hat{1}\right)
\Bigg]^{-1} \,.
\nonumber
\end{eqnarray}

$\check{G}_{d}(z)$ naturally embodies both the information about the quasi-ABS and that about the continuum spectrum of the JJ, respectively encoded in its singular and branch-cut part $\check G_{d}^{R}(z) = \check G^R(z) = \check G^{R}_{pol}(z) + \check G^{R}_{bcut}(z)$.

From the knowledge of the GF poles $z_{p}$, the polar GF can be rewritten in terms of a non-Hermitian Hamiltonian matrix $\check H_{eff} = \sum_p z_{p} (R_p \check Z^{-1}\check {P}_{p})$ and its residual matrix $\check Z = \sum_p R_p \check {P}_p$ \cite{Capecelatro2025}
\begin{equation}
\label{GR_eff_neq}
\check G^{R}_{pol}(z) = \quad
\check Z (z - \check H_{eff})^{-1} \,,
\end{equation}
where $R_p\check {P}_p$ is defined as 
\begin{equation}
R_p\check {P}_p=\frac{1}{2\pi i}\oint_{\mathcal{C}_p} \check G^R(z) dz 
\end{equation} 
with $\mathcal{C}_p$ a small contour around $z_p$. We have split the latter matrix in a rank-1 projector term satisfying $\check {P}_p^2 = \check {P}_p$ and $\mathrm{Tr} \check {P}_p = 1$ and a residual complex number $R_p$.

Interestingly, in the weak-coupling regime, that is for large superconducting gap $\Delta\gg\Gamma_{L/R}$, we can simplify the S lead self energy as $\check \Sigma_{S}=\Gamma\cos{\left(\frac{\phi}{2}\right)}\hat{\tau}_{x}\otimes \hat{1}+\gamma\sin{\left(\frac{\phi}{2}\right)}\hat{\tau}_{y}\otimes \hat{1}$, with $\check{Z}\rightarrow\check{1}$ and the GF singular part simply becoming $\check{G}_{pol}^{R}(z) = \left(z\check{1}-\check{H}_{eff}\right)^{-1}$.
This allows to immediately recognize the frequency-independent NH effective Hamiltonian in Eq.~\ref{Heff_NH} from Eq.~\ref{Gmatsu}.

\section{Josephson current formula}
\subsection{Josephson current in the infinite and finite \texorpdfstring{$\Delta$}{Delta} cases}
In Sec.~\ref{sec: 2_Model_Josephson_current} we introduced Eq.~\ref{Jpol_general} that allows to calculate the junction CPR starting from the quasi-ABS, i.e. the eigenvalues of $\check{H}_{eff}$, in every spectral configuration, e.g. ABS spectra hosting EPs or global zero-energy states (G-ZES) as in Fig.~\ref{fig: 1_EPs_in_ABS_spectrum}.

In Ref.\cite{Capecelatro2025}, to derive the above Andreev current formula, we started from the expression for free energy of the dressed dot in Matsubara formalism~\cite{Benjamin2007, Rozhkov1999}
\begin{equation}
    \mathcal{F}(\omega_{n})=-T\sum_{\omega_n}\ln\left(\det\left(\check{G}_{d}^{-1}(\omega_n)\right)\right)\,.
\end{equation}
Along this line, we can rewrite the free energy of the QD only for the singular part of the dot GF, $\check{G}^{R}_{pol}$, as a real-frequency integral~\cite{Benjamin2007, Rozhkov1999}
\begin{eqnarray}
	\label{free_energy}
    \mathcal{F}^{pol} &= \frac{1}{2\pi} \int_{-\infty}^{\infty}\mathrm{d}\omega\; n_F(\omega)   \mathrm{Im} \ln \det(\left(G^{R}_{pol}\right)^{-1})\\&= \frac{1}{2\pi} \int_{-\Delta}^{\Delta}  \mathrm{d}\omega\;  n_F(\omega) \mathrm{Im}\left[\prod_j\ln \left(\omega - z_j\right)\right]\nonumber\\ 
    &=\frac{1}{2\pi} \sum_j\int_{-\Delta}^{\Delta}  \mathrm{d}\omega\;  n_F(\omega) \mathrm{Im}\left[\ln \left(\omega - z_j\right)\right],\nonumber
\end{eqnarray}
accounting solely for the contribution of the GF poles, $z_j=\varepsilon_{j}-i\lambda_{j}$, i.e. the eigenvalues of $\check{H}_{eff}$. 
We remark that here, the summation upon all the complex poles has to be performed, regardless from the sign of their real part, since the levels broadening leads to finite occupation of the positive energy states also at zero temperatures.
By taking the phase derivative of Eq.~\ref{free_energy} one directly obtains Eq.\ref{Jpol_general}. 

The latter in the zero temperature limit, $T\rightarrow0$, becomes~\cite{Capecelatro2025}:
\begin{eqnarray}
	\label{Jpol_T0}
	J_{ABS}(\phi) &\overset{T\rightarrow0}{=}&  
        J_{\Re}+J_{\Im}=\sum_j  J_{\Re,j}+J_{\Im,j}=\nonumber\\&\overset{T\rightarrow0}{=}&\sum_j -\frac{e}{\pi}\partial_\phi \varepsilon_j\Big\{ \arctan\left(\varepsilon_j/\lambda_j\right) - \nonumber\\ 
        &&-\arctan\left((\Delta+\varepsilon_j)/\lambda_j\right)\Big\} +\\
		&& - \; \frac{e}{2\pi} \partial_\phi \lambda_j\ln\left( \frac{|z_j|^2}{(\Delta+\varepsilon_j)^2 +\lambda_{j}^2}\right)\,.
		\nonumber 
\end{eqnarray}

The above equation is the finite $\Delta$ counterpart of Eq.\ref{Jpol_T0_simp} and can be in principle applied also to different transport regimes, e.g. $\Delta\sim\Gamma_{L/R}$. 
We note that, contrary to the large gap limit $\Delta\gg\Gamma_{L/R}$, here Eq.~\ref{Jpol_T0} does not suffice in predicting the exact CPR since the contribution coming from the continuum part of the spectrum is also needed~\cite{Capecelatro2025}.

When taking the $\Delta\rightarrow\infty$ limit the leading order of the logarithmic factor in the imaginary dispersive terms that is 
\begin{equation}
    \ln\left( \frac{|z_j|^2}{(\Delta+\varepsilon_j)^2 +\lambda_{j}^2}\right)\sim\ln\left( \frac{|z_j|^2}{\Delta^2}\right)=2\ln|z_j|-2\ln\left(\Delta\right)\,
\end{equation}
might seem to show a divergent component proportional to $\ln(\Delta)$.
However, this term cancels when the summation over all the eigenvalues is performed. 

Indeed, we find that that 
\begin{equation}
	J_{\Im}(\phi) \overset{\Delta\rightarrow\infty}{\overset{T\rightarrow0}{=}}  
        -\frac{e}{\pi}\sum_j  \;  \partial_\phi \lambda_j\ln\left(|z_j|\right)\,,
\end{equation}
where we exploited the property that $\mathrm{Tr}\,{\Im\left(\check{H}_{eff}\right)}=-i4\Gamma_{N}$ as follows
\begin{equation}
\sum_j\partial_\phi \lambda_j\ln\left(\Delta\right)= \partial_\phi\sum_j \lambda_j\ln\left(\Delta\right)=-\partial_\phi \mathrm{Tr}\,{\Im\left(\check{H}_{eff}\right)}\ln\left(\Delta\right)=0\,,
\end{equation}
so that Eq.~\ref{Jpol_T0_simp} is obtained.

Eq.~\ref{Jpol_T0_simp} is found to be formally equivalent to the formula in Ref.~\cite{Shen2024},
\begin{equation}
    \label{Shen_formula}
    J\left(\phi\right)=-\frac{e}{\pi}\partial_\phi\Im\mathrm{Tr}\left(H_{eff}\ln H_{eff}\right)\,,
\end{equation}
when taking the effective Hamiltonian for the barrier degrees of freedom only, Eq.~\ref{Heff_NH}.
Indeed, by recalling that $\mathrm{Tr}\left(\check{H}_{eff}\ln \check{H}_{eff}\right)=\sum_j z_{j}\ln\left(z_{j}\right)$,  
Eq.~\ref{Shen_formula} reads
\begin{eqnarray}
    \label{Shen_formula_to_our_formula}
    J\left(\phi\right)&=&-\frac{e}{\pi}\partial_\phi\Im\sum_j z_{j}\ln\left(z_{j}\right)=\nonumber\\
    &=&-\frac{e}{\pi}\sum_j \Im \left(\partial_{\phi}z_{j}\right)\ln\left(z_{j}\right)-\frac{e}{\pi}\partial_{\phi}\sum_{j}\Im\left(z_{j}\right)=\nonumber\\
    &=&-\frac{e}{\pi}\sum_j \Im \left(\partial_{\phi}z_{j}\right)\ln\left(z_{j}\right)\,,
\end{eqnarray}
where we again used the fact that $\sum_{j}\Im z_{j}=-\sum_{j}\lambda_{j}=-i4\Gamma_{N}$.
Starting from Eq.~\ref{Shen_formula_to_our_formula},  Eq.~\ref{Jpol_T0_simp} can be retrieved by performing straightforward calculations.
\label{app: CPR_finite_infinite_Delta_cases}

\subsection{CPR in the time-reversal symmetric case, the contribution from the imaginary parts at the EPs}
\label{app: CPR_in_TRS_case}
Eq.~\ref{Jpol_T0_simp} that we report here for the sake of simplicity 
\begin{eqnarray}
	J_{ABS}(\phi) &\overset{\Delta\rightarrow\infty}{\overset{T\rightarrow0}{=}}&J_{\Re}+J_{\Im}= \sum_{j}J_{\Re, j}+\sum_{j}J_{\Im, j}\;, \;\; \mathrm{with} \nonumber\\
    J_{\Re, j}&\overset{\Delta\rightarrow\infty}{\overset{T\rightarrow0}{=}}&-\frac{e}{\pi} \partial_\phi \varepsilon_j\left( \arctan\left(\varepsilon_j/\lambda_j\right) - \frac{\pi}{2}\right),\\
	J_{\Im, j}&\overset{\Delta\rightarrow\infty}{\overset{T\rightarrow0}{=}}& -\frac{e}{\pi} \;\partial_\phi \lambda_j\ln\left( |z_j|\right)\,\nonumber.
\end{eqnarray}
accounts for the current coming from both real and imaginary parts phase dispersions for each quasi-ABS.

Contextually, the previous discussion in sec.~\ref{sec: 4_symmetries} provides a theoretical basis to understand why and when the imaginary part of the quasi-ABS is phase dependent. 
We are then able to trace a sharp difference between systems where the shifted Hamiltonian $\check{H}'$ has TRS and those with PHS$^\dagger$ only. In the first case, the imaginary current, $J_{\Im}$, is non-vanishing only within the regions between two ZE-EPs or beyond a FE-EP. In the second one, on the contrary, the imaginary part of the quasi-ABS contributes to the supercurrent also in the absence of EPs. As we will see in the following, Eq.~\ref{Jpol_T0_simp} declines in different ways in these two situations. 

In this subsection we analyze the current contributions in the presence of TRS on $\check{H}'$.

Despite not representing the general configuration for our setup, systems displaying the underlying TRS on $\check{H}'$ offer the opportunity to isolate the imaginary current contribution arising from the coalescence of two levels at the EPs~\cite{Pino2025}.

\begin{figure}
		\centering
		\includegraphics[scale=0.5]{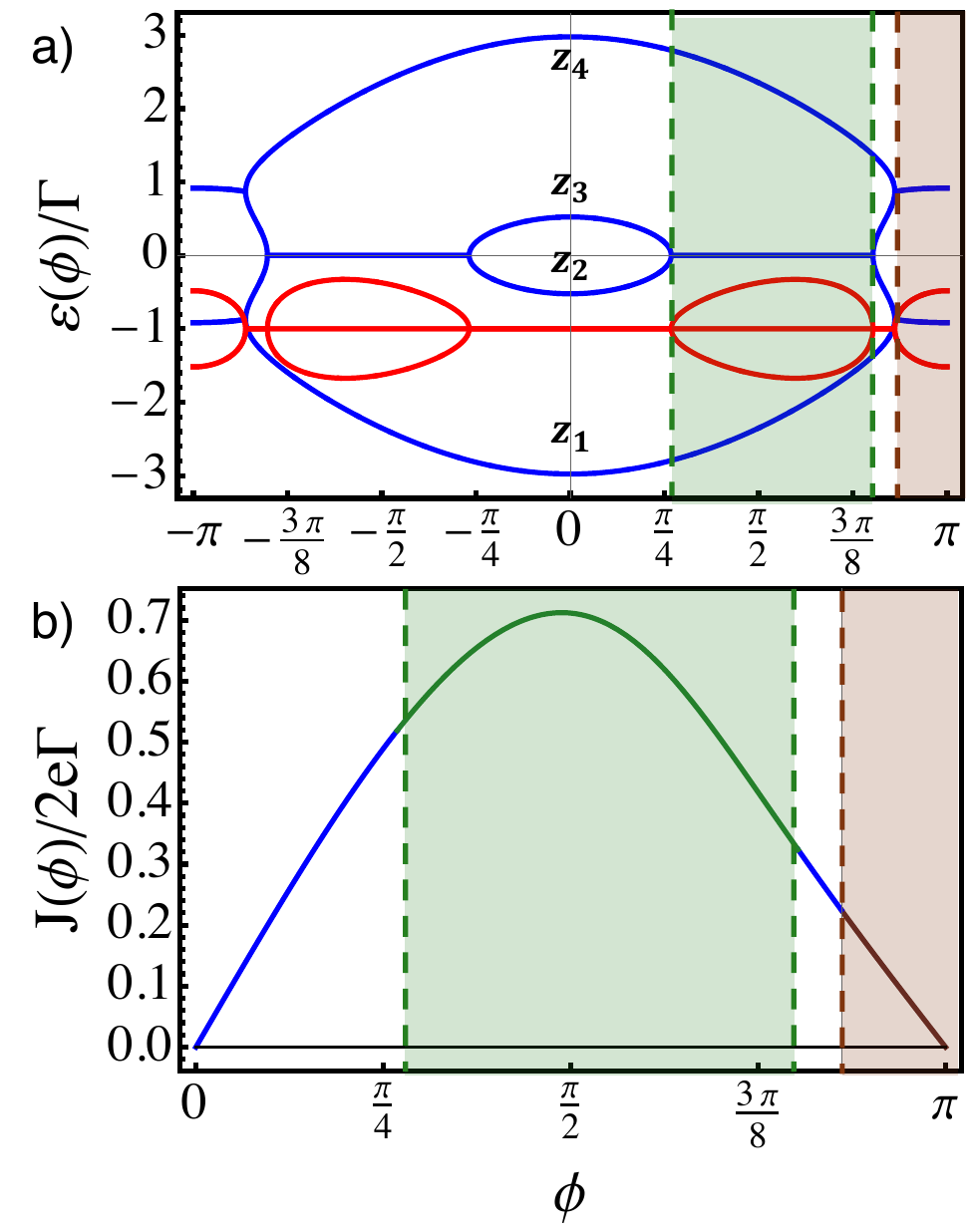}
		\caption{Example of Andreev levels spectrum (a) for a system with TRS and the corresponding CPR (b) computed piecewise by using the Eqs.~\ref{J_EPs}. The green shaded region between two ZE-EPs hosts dispersive ZES which contribute only to $J_{\Im}$. The brown shaded region between two FE-EPs (only half of it is shown) hosts a spectrum that is made of quartets $(z,-z,z^{*},-z^{*})$. 
        In the green shaded region the ZES only contribute to $J_{\Im}$ while in the brown shaded ones the levels coalescing at $\varepsilon\neq0$ are also dispersive in their real part, thus giving rise to $J_{\Re}$. The system parameters are $\Gamma=2,\,\gamma=0,\,B=1.1\,,\Gamma_{N}=1,\,\gamma_{N}=0.8\,,\theta=1$.}
		\label{fig: 3_example_ABS_CPR}
\end{figure}

In Fig.~\ref{fig: 3_example_ABS_CPR} (a), we show an example of the Andreev spectrum of such a junction hosting both ZE- and FE-EPs.
Here, the CPR can be thus computed piecewise as follows.
In regions corresponding to unbroken TRS phases for $\check{H}'$, the levels simply read $z_{j}=\varepsilon_{j}-i\lambda_{j}=\varepsilon_{j}-i\Gamma_{N}$. Here, $J_{\Im}=0$ and $J_{\Re}$ can be written only in terms of the quasi-ABS below the Fermi energy, e.g. $z_{1,2}$ in Fig.~\ref{fig: 3_example_ABS_CPR}: 
\begin{eqnarray}
\label{J_EPs}
    J_{\Re}&=&-\frac{2e}{\pi}\sum_{j=1,2}  \partial_\phi \varepsilon_j \arctan\left(\varepsilon_j/\lambda_j\right)=\\
    &=&-\frac{2e}{\pi}\sum_{j=1,2}  \partial_\phi \varepsilon_j \arctan\left(\varepsilon_j/\Gamma_{N}\right),\nonumber
\end{eqnarray}
where we exploit PHS$^\dagger$. Note that we recover the result of Ref.~\cite{Beenakker2024}, where the phase-independent broadening of the ABS is given by the average coupling to the metallic reservoir, $\Gamma_{N}$.

\begin{table*}[!t]
\begin{center}
\begin{tabular*}{\textwidth}{@{\extracolsep{\fill}} llcccll }
\toprule
\toprule
Parameters Regime & Symmetry name & Symmetry rule & Operator  & Square & Implied by & Class \\
 &  &  & in Nambu$\otimes$spin space &  &  &  \\
\midrule \midrule
Symmetries of $H$ & & & & & &\\
\midrule
any $\varepsilon_d\,, \ \Gamma_L \neq \Gamma_R$ 
  & PHS$^\dagger$ 
  & $\mathcal{T}_- H^* \mathcal{T}_-^{-1} = -H$
  & $\hat{\sigma}_y \hat{\tau}_y$ & $+1$ & & D$^\dagger$ \\
\midrule
any $\varepsilon_d\,, \ \Gamma_L = \Gamma_R$ 
  & TRS$^\dagger$ 
  & $\mathcal{C}_+ H^T\mathcal{C}_+^{-1} = H$
  & $\hat{1}$ & $+1$ & & BDI$^\dagger$ \\
  & CS 
  & $\Gamma_- H^{\dagger} \Gamma_-^{-1} = -H$
  & $\hat{\sigma}_y \hat{\tau}_y$ & $\setminus$ & PHS$^{\dagger}$ + TRS$^{\dagger}$ & \\
\midrule \midrule
Additional symmetries of $H'$  & & & & & &\\
\midrule
$\varepsilon_d = 0, \ \Gamma_L \neq \Gamma_R$ 
  & TRS 
  & $\mathcal{T}_+ H^* \mathcal{T}_+^{-1} = H$
  & $\hat{\tau}_x$ & $+1$ & & $\mathcal{S}_-$-AI \\
  & SLS 
  & $\Gamma_+ H \Gamma_+^{-1} = -H$
  & $\hat{\sigma}_y\hat{\tau}_z$ & $\setminus$ & PHS$^\dagger$ + TRS & \\
\midrule
$\varepsilon_d = 0, \ \Gamma_L = \Gamma_R$ 
  & TRS 
  & $\mathcal{T}_+ H^* \mathcal{T}_+^{-1} = H$
  & $\hat{\tau}_x$ & $+1$ & & $\mathcal{S}_{-+}$-CI\\
  & SLS 
  & $\Gamma_+ H \Gamma_+^{-1} = -H$
  & $\hat{\sigma}_y\hat{\tau}_z$ & $-1$ & PHS$^\dagger$ + TRS & \\
  & PHS 
  & $\mathcal{C}_- H^T \mathcal{C}_-^{-1}=-H$
  & $\hat{\sigma}_y\hat{\tau}_z$ & $-1$ & TRS$^\dagger$ + SLS & \\
\bottomrule
\end{tabular*}
\caption{
Zero-dimensional symmetry classification of the Hamiltonian $H_{eff}$ (table above) and of the shifted-reference $H'$ (table below).
For both tables each line inherits the symmetries of a more general case above it that has either $\varepsilon_d \neq 0$ or $\Gamma_L \neq \Gamma_R$, i.e. we don't write those symmetries again. $H'$ cases possess all symmetries of $H$ in the analogue cases, and we report only the additional ones when present.
Notice that setting $\varepsilon_d=0$ does not add any new symmetry to the complete Hamiltonian $H_{eff}$  while changing the symmetry class of the shifted Hamiltonian $H'$. TRS, PHS, CS and SLS indicate respectively time-reversal symmetry, particle-hole symmetry, chiral symmetry and sublattice symmetry. The \textbackslash~symbol is inserted whenever that information is not relevant for the classification.
}
\label{tab: Hsyms}
\end{center}
\end{table*}

Differently, green-shaded and brown shaded regions in Fig.~\ref{fig: 3_example_ABS_CPR} (a) correspond to broken TRS regions for $\check{H}'$, with the former enclosed by two ZE-EPs and the latter occurring after a FE-EP. 
These phase intervals are characterized by a non-vanishing $J_{\Im}$ arising from the levels that coalesce together in the real part and bifurcate on the imaginary axes
\begin{eqnarray}
\label{J_EPs}
   J_{\Im}^{\mathrm{ZE-EPs}}&=&-\frac{e}{\pi}\partial_{\phi}\lambda_{2}\log\left(\frac{|\lambda_{2}|}{|\lambda_3|}\right) \,\,\,\mathrm{for\,\,ZE-EPs}\,,\\
    J_{\Im}^{\mathrm{FE-EPs}}&=&-\frac{e}{\pi}\partial_{\phi}\lambda_{1}\log\left(\frac{|z_{1}|}{|z_2|}\right) \,\,\,\mathrm{for\,\,FE-EPs}\,,
\end{eqnarray}
where $z_{2,3}=\varepsilon_{2,3}-i\lambda_{2,3}$ and $z_{1,2}=\varepsilon_{1,2}-i\lambda_{1,2}$
are the pair of levels respectively involved in the ZE-EPs and in the FE-EP, Fig.~\ref{fig: 3_example_ABS_CPR}. Here, Eq.~\ref{J_EPs} is the result of Ref.~\cite{Pino2025} when considering solely the poles of the barrier GF~\cite{Capecelatro2025}.
Note that in the brown-shaded interval, where the spectrum is made of quartets $(z,z^{*},-z,-z^{*})$, the coalescing levels feature a phase-dispersive behavior also in the real part, with a finite $J_{\Re}$ component,
\begin{equation}
    \label{JRe_FE_EP}
    J_{\Re}^{\mathrm{FE-EPs}}=-\frac{e}{\pi} \sum_{j=1,2}\partial_\phi \varepsilon_j\left( \arctan\left(\varepsilon_j/\lambda_j\right) - \frac{\pi}{2}\right)\,.
\end{equation}

This analysis of the CPR components by intervals is feasible only when the spectrum of $\check{H}'$ is symmetric with respect to both real and imaginary axis, i.e. in the presence of an underlying TRS, Fig.~\ref{fig: 8_ABS_spectrum_epsdne0_epsdequal0} (a).

When TRS lacks, there is no constraint upon $\lambda_{j}(\phi)$, thus the imaginary part of quasi-ABS features a phase-dispersive background that sums to the EP current and cannot be simply traced out. 
Furthermore, formulas analogous to Eq.~\ref{J_EPs} cannot be derived, since the two levels coalescing at the EPs bifurcate with different derivatives and therefore do not sum into a single term.
This difference is compensated by the phase dispersion of the imaginary parts of the remaining levels, see Appendix.
Finally, unlike the case with TRS on $\check H'$, in general all levels contribute to the CPR, pointing out a crucial difference from the Hermitian case, where only the ABS below the Fermi energy are involved.

In the main text the results are computed by the means of Eq.~\ref{Jpol_T0_simp} that holds also for $\varepsilon_d\neq0$.

\section{Symmetries of the NH Hamiltonian}
\label{app: symmetries}
The system has been shown to host both zero-energy
and finite-energy EPs~\cite{Solow2025} (ZE-EPs and FE-EPs) which are found to be respectively fragile and robust under an additional gate voltage tuning the dot energy $\varepsilon_{d}$. The type and robustness of the EPs are determined by the symmetries of $\check{H}_{eff}$. Further, these encode the information about the phase dispersion of the complex Andreev spectrum, as we show in sec.~\ref{sec: 4_symmetries}.

In this Appendix, we report all 
symmetries classes to which the system belongs depending on the choice of the junction parameters \cite{Kawabata2019, Gong_2020}. 
The symmetry classes together with the corresponding symmetry operators are listed in Tab.~\ref{tab: Hsyms}.
Here, the both the complete NH Hamiltonian, $\check{H}_{eff}$, and its shifted version, $\check{H}'$, are considered as a zero-dimensional system (evaluated at a fixed generic phases bias), which is sufficient to characterize the 
quasi-ABS spectrum.

In the whole parameter space, the effective Hamiltonian has always the adjoint particle-hole symmetry (PHS$^\dagger$), imposed by the Nambu doubling.
As a consequence it exists an operator $\mathcal{T}_- \mathcal{K}$, which anticommutes with the Hamiltonian, for which, given an energy eigenvector $\ket{e_j}$ with eigenvalue $z_j=\varepsilon_j-i\lambda_j$, $\mathcal{T_-}\ket{e_j}$ is also an energy eigenvector but with with eigenvalue $-z_j^*$.
This condition extends to non-Hermitian matrices the concept of particle-hole symmetry (PHS) while preserving the causality constraint of the retarded GF, $G^{R}_{pol}$, from which $\check{H}_{eff}$ is extracted \cite{Capecelatro2025}. 
PHS$^{\dagger}$ can be then fulfilled both by pairs $(z_j,-z_j^*)$
or by unpaired levels with $\Re{z_j}=0$.
At ZE-EPs, when a pair of particle hole symmetric levels $(z=\varepsilon-i\lambda,-z^{*}=-\varepsilon-i\lambda)$ coalesces on the real axis, $\Re\{z\}=0$, a transition between different PHS$^{\dagger}$ phases occurs within the corresponding two-level subspace and, after the EP, the pair disassembles into two unpaired levels, $(i\lambda_{+}, i\lambda_{-})$.

More generally, for anti-unitary symmetries such as PHS$^{\dagger}$, a transition in the eigenvalues is necessarily accompanied by a concomitant transition in the associated eigenstates~\cite{Zha23}.
On the first side of the ZE-EP, the two eigenstates are mapped onto the other by the symmetry operator whereas, on the other side, they are independent eigenvectors.
This argument, however, does not apply to the FE-EPs. Indeed, the eigenvalues (and the eigenvectors) involved in this kind of EPs are not related by PHS$^{\dagger}$ at all.
Although their emergence may be explained by restricting the Hamiltonian to the relevant subspace~\cite{Solow2025}, 
this does not come out as a consequence of any symmetry of $\check{H}_{eff}$ (we list them in Tab.~\ref{tab: Hsyms} of Appendix~\ref{app: symmetries} for completeness). 
In other words, FE-EPs appear as simple accidental features.
Their emergence is fully explained when looking at the symmetries of the shifted Hamiltonian $\check{H}'$. Only the TRS on $\check{H}'$ present at $\varepsilon_{d}=0$ stabilizes the FE-EPs, which are indeed fragile with respect to an external gate potential applied on the dot~\cite{Solow2025}, as it is explained in sec.~\ref{sec: 4_symmetries}.

Thus, in the resonant regime, the structure of the spectrum is shaped by PHS$^{\dagger}$ on $\check{H}_{eff}$ and a TRS on $\check{H}'$.
This symmetry implies that the eigenvalues of $\check{H}_{eff}$ appear in pairs $(z_j-i \Gamma_N,z_j^* -i \Gamma_N)$ or with unpaired levels with $\Im\{z_j\} = -i \Gamma_N$, corresponding respectively to \emph{unbroken} and \emph{broken} phases. The transition between the two can happen either at zero energy at ZE-EPs or at finite energy at FE-EPs.
In this limit the imaginary part shows a phase dispersion only in the regions where this TRS is broken. 
Further, we observe that the \emph{quartet} regime in Sec.~\ref{sec: 3_spectral_regimes_GZES}, with no ZES and a non-zero $J_{\Im}$ contribution, is only possible due to the combination of PHS$^{\dagger}$ and TRS and it cannot be achieved off-resonance.

We here stress that in the most general situation TRS 
is not present and emerges only at $\varepsilon_{d}=0$ for asymptotically large $\Delta$. Indeed, we note that the full self-energy term of S leads in Eq.~\ref{S_leads_self_energy_Delta_finite} does not fulfill TRS, so that other transport regimes, e.g. the short JJ limit achieved for $\Delta\ll\Gamma$, are characterized by quasi-ABS phase-dispersive imaginary part also in the absence of EPs~\cite{Capecelatro2025}.

\end{document}